\documentclass[useAMS,usenatbib,usegraphicx,useamssymb]{mn2e}
\usepackage{graphicx}
\usepackage{amssymb}
\usepackage{floatpag}
\usepackage{float}
\usepackage[section]{placeins}
\usepackage{relsize}
\usepackage{amsmath}
\usepackage{url}
\usepackage{longtable}
\usepackage{supertabular}

\def \apj {ApJ}
\def \apjs {ApJS}
\def \apjl {ApJL}
\def \mnras {MNRAS}
\def \aap {A\&A}

\def \araa {ARAA}

\def \nat {Nature}

\def \nar {New. A. Rev.}
\def \apss {Ap\&SS}
\def \cjaa {Chin. J. Astron. Astrophys.}

\title[]{A redshift -- observation-time relation for gamma-ray bursts: evidence of a distinct sub-luminous population}

\author[]{E. J. Howell$^{1}$\thanks{E-mail:eric.howell@uwa.edu.au} and D. M. Coward$^{1}$\\
$^{1}$School of Physics, University of Western Australia, Crawley WA 6009, Australia\\}

\begin{document}
\linespread{1.0}

\maketitle

\begin{abstract}
We show how the redshift and peak-flux distributions of gamma-ray bursts (GRBs) have an observation-time dependence that can be used to discriminate between different burst populations.
We demonstrate how observation-time relations can be derived from the standard integral distributions and that they can differentiate between GRB populations detected by both the BATSE and \emph{Swift} satellites. Using \emph{Swift} data we show that a redshift--observation-time relation (log\,$Z$\,--\,log\,$T$) is consistent with both a peak-flux\,--\,observation-time relation (log\,$P$\,--\,log\,$T$) and a standard log\,$N$\,--\,log\,$P$ brightness distribution. As the method depends only on rarer small-$z$ events, it is invariant to high-$z$ selection effects. We use the log\,$Z$\,--\,log\,$T$ relation to show that sub-luminous GRBs are a distinct population occurring at a higher rate of order $150^{+180}_{-90} \mathrm{Gpc}^{-3}\mathrm{yr}^{-1}$. Our analysis suggests that GRB 060505 -- a relatively nearby GRB observed without any associated supernova -- is consistent with a sub-luminous population of bursts. Finally, we show that our relations can be used as a consistency test for some of the proposed GRB spectral energy correlations.

\end{abstract}

\begin{keywords}
gamma-rays: bursts -- gamma-ray: observations -- methods:data analysis -- supernovae: general -- cosmology: miscellaneous

\end{keywords}

\section{Introduction}

Multi-wavelength observations have shown that $\gamma$-ray bursts (GRBs) are the most luminous\footnote{In terms of electromagnetic radiation per unit solid angle.} and distant transient events in the Universe \citep{Greiner_furthestGRB6pt7_08,2011arXiv1105.4915C}. GRBs have been generally categorized into two populations: spectrally soft long duration bursts related to core-collapse events (LGRBs/Type II); harder short duration bursts possibly resulting from compact star mergers (SGRB/Type I).

In addition to these two main populations of bursts it has been suggested there exist two sub-populations: sub-luminous GRBs (SL-GRBs) and SGRBs with extended emissions (SGRB-EE). SL-GRBs are of the long-duration type and have isotropic equivalent $\gamma$-ray luminosities 2-3 orders of magnitude below classical LGRBs \citep{coward_LLGRB_05,murase_06,GuettaDellaValle_2007,Imerito_08}. The lower energy emissions mean they are only detected at low-$z$ -- as such, four of the six GRBs with unambiguous spectroscopically confirmed GRB-supernova associations were from this category. SGRB-EEs emissions have been given a separate classification in the second Swift catalogue \citep{sakamoto_BAT2_2011}. These bursts show an initial SGRB like short hard spike ( $< 2$\,s) followed by a faint softer emission ($\gtrsim 100$\,s) \citep{norris_06, norris_2011, page_SGRB_EE_2006, Perley_extendedSGRB_08, zhang_openQs_2011}. \\
\indent There is still no clear consensus that these sub-categories arise from different progenitor systems or are simply rarer events from the tail of the respective short/long burst distributions. Attempts to address this have generally been based on statistical arguments \citep{sodoburg_06_LLGRBRate_06,GuettaDellaValle_2007}, fits to the log\,$N$--\,log\,$P$, peak flux, or `brightness distribution' of bursts \citep{Pian_LLGRBs_06} or through simulation \citep{coward_LLGRB_05,Virgilii_LLGRBs_08}. \\
\indent The goal of this paper is to demonstrate an alternative strategy using the relative time records of the bursts. This approach exploits the fact that different astrophysical transient populations will have different local rate densities. We show that by recording the arrival times of the rarest events in a time series e.g. the closest or brightest of a cosmological population, one can produce a rate dependent data set with a unique statistical signature \citep{coward_PEH_05}. By constraining the data using an observation-time dependent model that is highly sensitive to the rate density, we demonstrate how this alternative approach can untangle different source populations. In this study we will specifically address the issue of whether SL-GRBs are a distinct population of GRBs. The outline of the paper is as follows:

In section 2 we present an overview of GRB population studies. Section 3 will set the scene in regards to the observation-time dependence of transient events and Section 4 will describe the data extraction methodologies used in this study. A standard theoretical framework will follow in Section 5.

Section 6 describes observation-time dependent models for both peak flux (log\,$P$--\,log $T$) and redshift (log\,$Z$--\,log\,$T$) showing how they follow seamlessly from the relative integral distributions of transient sources. In Section 7, parameters for both the BATSE\,\footnote{The Burst and Transient Source Experiment (BATSE) on the Compton Gamma-Ray Observatory was launched in 1991, and recorded over 2000 GRBs at a rate of around one a day for around 9 years of operation.} and \emph{Swift} LGRB populations will be obtained using a standard differential log\,$N$--\,log\,$P$ distribution. These parameters will be used in Section 8 to constrain the peak-flux -- observation-time data from both detectors using a log\,$N$--\,log\,$P$ model. Doing so demonstrates that the method is both consistent with a standard brightness distribution and is detector independent.

Section 9 will demonstrate the use of the previously derived parameters in the redshift -- observation-time domain to constrain \emph{Swift } data using the log\,$Z$--\,log\,$T$ model. In Section 10 we apply our methods to the \emph{Swift} SL-GRB population to further demonstrate how the method can be used as a tool to discriminate between different source populations. We show that the method uses only the closest or brightest of a population; thus many of the selection biases that plague GRB observations can be bypassed. We discuss our findings and present our conclusions in Section 11.

\section{Gamma-ray Burst populations}
\label{section_GRB_populations}

The catagorisation of GRBs was traditionally based on the bimodal distribution of $T_{90}$ durations observed by BATSE \footnote{The $T_{90}$ duration is the time during which the cumulative counts increase from 5\% to 95\% above background.} \citep{Kouveliotou_1993} and their hardness ratio in the spectral domain. These criteria separated GRBs into hard SGRBs ($T_{90}<2\,\mathrm{s}$;\,hard spectra) and softer LGRBs ($T_{90}\geq 2\,\mathrm{s}$;\,soft spectra).

Electromagnetic observations of LGRBs and SGRBs have provided strong evidence for different progenitors. LGRBs have been associated with the deaths of massive stars \citep{WB_06,Hjorth_LGRB_SN,Stanek_LGRB_SN} and have subsequently been found in or near dense regions of active star-formation, predominantly dwarf starburst field galaxies \citep{Fruchter_grb_gal}. For SGRBs, which have contributed around 25\% and 10\% of the BATSE and \emph{Swift} GRB samples respectively \citep{Guetta_SGRB_GW_08}, the leading progenitor model is the merger of compact neutron stars and/or black hole binaries. The association of an older stellar population with these bursts is supported by their occurrence in both early and late-type galaxies, as well as field and cluster galaxies.\\
\indent There exist, however, a number of ambiguities in the categorisation schemes of GRB populations. LGRBs, such as GRB 060614 and GRB 060505, showed no evidence of a supernova, despite extensive follow up campaigns \citep{Zhang_GRBCLassification_06}. Additionally, it has been suggested that the two closest recorded bursts, GRB 980425 (36 Mpc) and GRB 060218 (145 Mpc), along with GRB 031203, associated with a host galaxy at $\sim 480$ Mpc \citep{feng_09} and GRB 100316D \citep{Starling_2011}($z\sim 0.06$) make up a sub-class of SL-GRBs \citep{cobb_06,Liang_07,GuettaDellaValle_2007,Virgili_09}. This class of GRB have isotropic equivalent $\gamma$-ray energy emissions typically several orders of magnitude below those of standard long-duration GRBs  \citep{murase_06,GuettaDellaValle_2007,Imerito_08} suggesting that they could form a unique population of bursts that due to their relatively close proximity, must be occurring at a higher rate. The suggestion that SL-GRBs could be just normal LGRBs viewed off-axes has also been considered. This scenario was discounted by \citet{Daigne_2007} on statistical arguments as it would: a) result in a far higher local rate density than expected from LGRBs ; b) require much narrower opening angles for LGRBs than typically derived from the breaks in afterglow lightcurves.\\
\indent While the LGRB/supernovae connection is firm for the majority of these bursts, the progenitors behind SGRBs, which are rarer and more difficult to localise, are less certain \citep{meszaros_RPP_2006,Nakar_SGRBs_Rev_07}. Around 20\% of the SGRBs detected by \emph{Swift} have been followed by an extended emission lasting up to 100\hspace{1mm}s \citep{norris_06,Perley_extendedSGRB_08} leading to suggestions that different progenitors produce these bursts \citep{norris_2011}. Candidate systems include the birth of a rapidly rotating proto-magnetar produced through NS-NS merger or the accretion induced collapse of a white-dwarf \citep{Metzger_2008,Bucciantini_2011}. \citet{chapman_09} further postulated that an initial spike produced by a soft $\gamma$-ray repeater (SGR) in a galaxy of close proximity could mimic a SGRB and found that a dual population luminosity function based on both SGR giant flare properties and SGRB luminosities was consistent with BATSE data. Interestingly, \citet{Vavrek_MNRAS2008} found the sky distribution of SGRBs to be anisotropic, implying a fraction at close proximity. Certainly, this suggests that classification of shorter duration bursts could be detector dependent \citep{page_SGRB_EE_2006}.\\
\indent The various ambiguities have motivated a number of authors to re-define different classes of GRB through a number of properties including: spectral features, associated supernova, stellar population, host galaxy, location in the host galaxy and progenitor \citep{Zhang_GRBclass_06,Bloom_2008}. This has led to the Type I (compact object mergers) and Type II (core-collapse) scheme.\\
\indent The existence of a third  population of bursts with intermediate duration has also been suggested. The first evidence of this additional population was provided though statistical analysis of the BATSE distribution \citep{Mukherjee_ApJ1998,HorvathApJ1998}. This was later supported by analysis using data from the \emph{BeppoSAX} \citep{Horvath_ApSS2009}, \emph{RHESSI} \citep{Ripa_AA2009} and \emph{Swift} satellites \citep{Horvath_ApJ2010ApJ}. To formulate a physical model for an intermediate class of bursts, \citet{Veres_2010} showed that intermediate bursts have a lower than average peak flux distribution and suggested that this group, which are spectrally softer than LGRBs, may be related to X-ray flashs (XRFs) - these events have a dominant fraction of the total prompt fluence detected in X-ray rather than $\gamma$-ray.

%
%
%

\section{Exploiting the time dimension}
\label{section_time_dimension}

In this study we show that the time-record of GRB observations can be used as a tool to untangle different GRB populations. This largely overlooked quantity features strongly when one considers the observation of the \emph{rarest} events in a population i.e. those events from the tail of the distribution which occur at low-$z$ or have exceptional brightness in comparison with the average. \cite{coward_PEH_05} showed that these events posses a unique rate dependent statistical signature that can be described by the `probability event horizon' (PEH) concept. By recording successively rarer events as a function of observation-time, a data set can be produced and constrained by a rate dependent model \citep{coward05b,howell07}.

\begin{figure}
  \includegraphics[scale = 0.55,bbllx = 100pt,bblly =290pt, bburx = 500 pt, bbury = 600 pt,origin=lr]{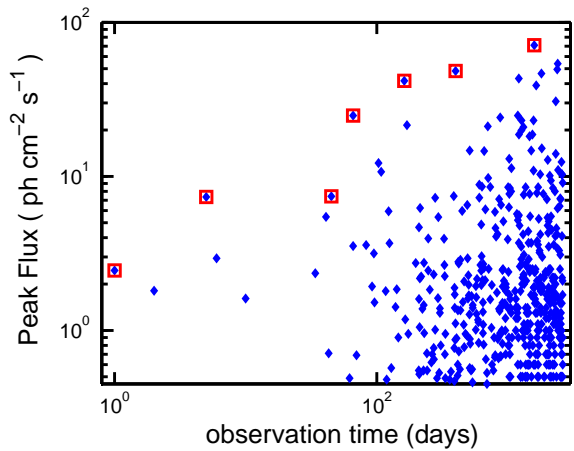}\\
  \caption{The \emph{Swift} LGRB peak flux data plotted against observation-time. The plot illustrates how the probability of observing a bright event increases with observation-time. Successively brighter events - termed \emph{PEH data} -  are indicated by squares.}
  \label{swift_logPlogT_data}
\end{figure}

The basic concept is illustrated in Figure \ref{swift_logPlogT_data} showing the \emph{Swift} LGRB peak flux data plotted against observation-time -- we will often refer to such a time-series as $P(T)$ data (or $Z(T)$ in the redshift domain). Using a log-log plot it is apparent that successively brighter events have an observation-time dependence -- the longer you observe, the greater the probability of observing an exceptionally bright event. Successively brighter events are indicated by squares -- these are termed \emph{PEH} data.

To constrain PEH data \citet{howell07_APJL} extended the standard Euclidean derivation of a log\,$N$--\,log\,$P$ brightness distribution to show how including a time dependence could produce a peak flux--observation-time power law relation, log\,$P$--\,log\,$T$. They also illustrated how some processing could improve the amount of PEH data. In an application to \emph{Swift} LGRB data, they demonstrated that the relation could be used as a probe of the event rate density of the sources. Additionally, the time dependence allowed the method to be used as a tool to predict the likelihood of future events.

In this paper we extend the previous study to present a cosmological log\,$P$--\,log\,$T$ relation and show that the principle also extends to the redshift domain resulting in a log\,$Z$--\,log\,$T$ relation. For the latter case, the PEH data is a measure of the geometrical distribution of the source population. A log\,$Z$--\,log\,$T$ curve can be thought of as a horizon, defined by successively closer events, that approach a central observer as a function of observation-time \citep{coward_PEH_05}. As the successively closer events approach the local low-$z$ regime rapidly, the GRB selection function \citep{Coward2007NewAR} and high-$z$ selection effects such as the `redshift desert' \citep{coward_GRBDessert_08} have a negligible effect.

In the next section we will describe the data extraction principles that will be employed in the latter part of this study.

\section{Data extraction methodology}
\label{section_data_extraction}

To extract PEH data, we follow the \emph{FromMax} method used by \citet{howell07_APJL} to constrain the \emph{Swift} LGRB sample. This invoked the temporal cosmological principle: for time scales that are short compared to the age of the Universe, there is nothing special about the time we switch on our detector. Therefore, the $P(T)$ time-series can be treated as closed loop, i.e. the start and endpoints of the time series can be joined and the start time set immediately after the brightest event. Successively brighter events and their observation-times are then recorded to produce a PEH data set of peak fluxes.

\begin{figure}
  \includegraphics[scale = 0.42,bbllx = 20pt,bblly =50pt, bburx = 500 pt, bbury = 530 pt,origin=lr]{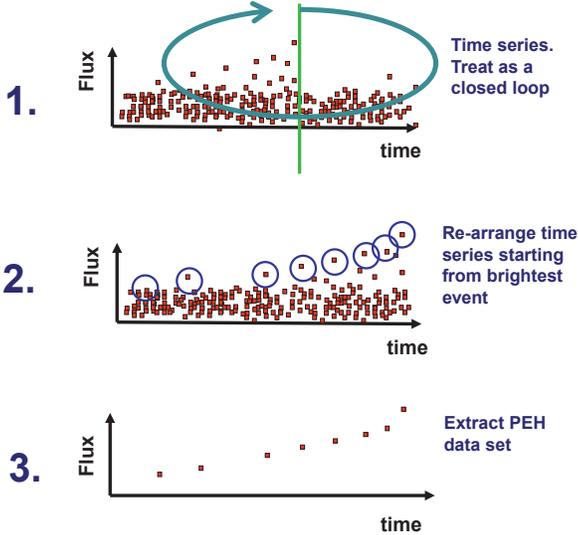}\\
  \caption{A cartoon illustrating the FromMAX data extraction principles of \citet{howell07_APJL}. Treating the data as a closed loop one can select a start point from which a data set of successively brighter events can be extracted. This data set is called PEH data.}
  \label{peh_cartoon}
\end{figure}

Figure \ref{peh_cartoon} illustrates the concept. Using this procedure \citet{howell07_APJL} showed that PEH data set can be obtained through both temporal directions -- for simplicity we will only record data in the forward direction.


Employing this technique circumvents the possibility of a loud event occurring early in a time series; this would minimize the amount of output data as the next largest event would most probably occur near the end of the time series. Such a situation could be encountered by a detector operating with a high energy cutoff -- a bias could be introduced producing a large number of events reaching the threshold. The advantage will become apparent in section \ref{BATSE logPlogT data} through analysis of the BATSE sample of bursts which contained a number of bright events at early observation-times.

Another feature of the FromMax method is that it ensures the total time duration of the PEH output is always equivalent to that of the total observation-time -- this ensures a well ordered data sample is produced with a consistent time signature. We will show later through 2D Kolmogorov Smirnov testing that the improved data set retains the statistical signature of the original.


%
%
%
%

To apply the procedure to a sample of $P(T)$ data we first define the brightest event by $P_*$ with an observation-time stamp $T_*$ and denote the time of the last, most recent occurring event, as $T_\mathrm{max}$. Treating the data as a closed loop we reorder the data starting from the first event after $ P_*$. The time stamps of the re-ordered data set $P^{\backprime}(T_{P}^{\backprime})$ are now defined as:
\vspace{-2mm}
\begin{equation}\label{peh_time}
  T^{\backprime}_{P}=\biggl \lbrace{ \begin{array}{lll}
                   T-T_* & T>T_* \\
                   \\
                   T+T_{\mathrm{max}}-T_* & T\leq T_*
                 \end{array}}
\end{equation}



To obtain a PEH data set we start to extract our data from the first minimum $P_{\mathrm{min}}^{\backprime}=P_{i}^{\backprime} < P_{i+1}^{\backprime}$ -- this additional step is to minimise the effect of an early bright event. A PEH data set is then obtained by recording successively brighter events $(P_{i}^{\backprime},T_{P,i}^{\backprime})$ satisfying the condition $P_{i+1}^{\backprime}>P_{i}^{\backprime}$ for $P_{i}^{\backprime}\geq P_{\mathrm{min}}^{\backprime}$.

To determine the PEH data set in the redshift domain one applies similar principles, treating the data as a closed loop but re-ordering the data from the first event after the closest redshift event $Z_0$. The timestamps for the re-ordered set $Z^{\backprime}(T^{\backprime}_{Z})$ are then given by:

\begin{equation}\label{peh_time}
  T_{Z}^{\backprime}=\biggl \lbrace{ \begin{array}{lll}
                   T-T_0 & T>T_0 \\
                   \\
                   T+T_{\mathrm{max}}-T_0 & T\leq T_0
                 \end{array}}
\end{equation}


\noindent A PEH data set is obtained by extracting data from the first maximum $Z_{\mathrm{max}}^{\backprime}=Z_{i}^{\backprime} > Z_{i+1}^{\backprime}$, recording successively closer events $(Z_{i}^{\backprime},T_{Z,i}^{\backprime})$ satisfying the condition $Z_{i+1}^{\backprime}<Z{i}^{\backprime}$ for $Z_{i}^{\backprime}\geq Z_{\mathrm{max}}^{\backprime}$.

\section{Theoretical Framework}
\label{section_theoretical_framework}

\subsection{GRB Flux and luminosity relations}

We firstly define an isotropic equivalent photon luminosity in the source frame as:
\vspace{-0.5mm}
\begin{equation}\label{eq_rest_frame_luminosity}
    L = \int^{10000\, \mathrm{keV}}_{1 \mathrm{keV}} S(E)\mathrm{d}E\,,
\end{equation}

\noindent where $S(E)$ is the differential rest-frame source luminosity (in ph s$^{-1}$ keV$^{-1}$). To define the observed peak photon flux (photons per cm$^{2}$ per second) observed within a detector band $E_{\mathrm{min}}\hspace{-2.0mm}<\hspace{-2.0mm}E\hspace{-2.0mm}<\hspace{-2.0mm}E_{\mathrm{max}}$ and emitted by an isotropically radiating source at redshift $z$ one must
perform two modifications. Firstly, the observed photon flux is modified to account for the missing fraction of the gamma ray energy seen in the detector band $b = \int^{E2}_{E1}S(E)dE / \int^{10000}_{1}S(E)dE $. Secondly a cosmological $k$-correction,
$k(z) =  \int^{E2}_{E1}S(E)dE / \int^{E2(1 + z)}_{E1(1 + z)}S(E)dE $ is applied. With these two modifications the standard definition for flux becomes:
\vspace{-0.5mm}
\begin{equation}\label{eq_rest_frame_luminosity}
    P =  (1 + z) \frac{L}{4 \pi d_{\mathrm{L}}(z)^{2} } \frac{b}{k(z)} \,.
\end{equation}

\noindent Substituting in for $b$ and $k(z)$ one obtains the familiar relation:
\vspace{-0.5mm}
\begin{equation}\label{eq_peakflux}
    P = \frac{ (1 + z) \int^ {(1 + z)E_{\mathrm{max}}}_{(1 + z)E_{\mathrm{min}}} S(E)\mathrm{d}E}{4\pi d_{\mathrm{L}}(z)^{2}}\,,
\end{equation}

\noindent where $d_{\mathrm{L}}(z)$ is the luminosity distance. The $(1 + z)$ factor is included as the standard definition of $d_{\mathrm{L}}(z)$ is valid for an energy flux, but $P$ here is given as a photon flux \citep{Meszaros2011}. For long duration GRBs the function $S(E)$ is typically modeled by a Band function \citep{Band03} which we use with high and low energy spectral indices of -2.25 and -1 and a break energy of 511 keV.

%
%


\subsection{GRB Luminosity Function}

A number of different forms have been suggested for the LGRB Luminosity Function (LF). To minimise our free parameters we the single power law form used by \citet{pm01} which has an exponential cutoff at low luminosity:
\vspace{-0.5mm}
\begin{equation}
\Phi(L) = \Phi_{0}\hspace{1.0mm}\left(\frac{L}{L_{*}}\right) ^{-\alpha} \mathrm{exp}\left( -\frac{L_{*}}{L}\right)\,.
\end{equation}

\noindent Here, $L$ is the peak rest frame photon luminosity in the 1-10000\,keV energy range, $\alpha$ ensures an asymptotic slope at the bright end and $L_{*}$ is a characteristic cutoff scaling. The normalisation coefficient for this LF is given by $\Phi_{0}=[L_{*}\Gamma (\alpha - 1)]^{-1}$. Based on the studies of \citet{Meszaros_ApJ_1996ApJ, Meszaros_ApJ_1995ApJ, Reichart_ApJ_1997,Butler2010} we will assume no luminosity evolution with redshift.

\subsection{GRB rate evolution}

To model GRB rate evolution, $R_{\rm GRB}(z)$, we assume that LGRBs track the star formation history of the Universe \citep{Meszaros_AA_2006} and normalise a star formation rate model, $R_{\rm SF}(z)$, (in units of mass converted to stars per unit time and volume) to the local $(z = 0)$ rate. Multiplying by the local rate density $\rho_{0}$ (in $\mathrm{Gpc}^{-3}\mathrm{yr}^{-1}$) allows one to extrapolate rate evolution to cosmological volumes:
\vspace{-0.5mm}
\begin{equation}\label{eq_sfr}
    R_{\mathrm{GRB}}(z)= \rho_{0} \frac{R_{\rm SF}(z)}{R_{\rm SF}(z=0)}\,.
\end{equation}

\noindent For $R_{\rm SF}(z)$ we use the model of \citet{Li_Metallicity_08}, which was obtained by adding ultraviolet and infared measurements to the sample of \cite{hb_sfr_06}. This takes the form:
\vspace{-0.5mm}
\begin{equation}\label{eq_sfr}
  R_{\rm SF}(z) = \frac{(0.02+0.12 z)}{1+(z/3.23)^{4.66})}\,.
  \end{equation}

A number of studies suggested that cosmic metallicity evolution must also be considered in any rate evolution model for LGRBs \citep{Li_Metallicity_08,Modjak_Metallicity_08}. A metallicity dependence results from the requirement that Wolf-Rayet (WR) stars should retain sufficient rotation to power a GRB -- therefore angular momentum losses through stellar-wind induced mass-loss must be minimized \citep{Woosley_GRB_Progenitors_06}. As wind-driven mass loss in WR stars is understood to be dependent on a high enough fraction of iron, a low-metallicity environment is an essential requirement \citep{vink_WR_05, WoosleyJankaNature_05}.

Recent results from \citet{Elliott2012}, driven by observations of LGRBs in metal rich galaxies, suggest that interpretations of luminosity and metallicity evolution could simply result from a misunderstanding of various redshift biases \citep{coward_GRBDessert_08}. As we will demonstrate later in section \ref{section_logZlogT_results}, the methods employed in this paper are sensitive to the rarer low-$z$ population of bursts. Therefore, for clarity, and in support of the uncertainties discussed above, we will ignore the effects of metallicity evolution in this study.



\subsection{The all sky event rate equation of GRBs}

The number of GRBs per unit time within the redshift shell $z$ to $z + \mathrm{d}z$ with luminosity $L$ to $L + \mathrm{d}L$ is given by:
\vspace{-0.5mm}
\begin{equation}
\label{dN}
\frac{\mathrm{d}N}{\mathrm{d}t \mathrm{d}z \mathrm{d}L }\hspace{0.5mm} = \psi(z)
 \frac{\mathrm{d}V (z) }{\mathrm{d}z } \frac{ R_{\mathrm{GRB}}(z) }{( 1 + z)}\,\mathrm{d}z{\hspace{0.5mm}}\Phi(L)\,.
\vspace{0mm}
\end{equation}

\noindent Here the $(1 + z)$ factor accounts for the time dilation of the observed rate by cosmic expansion; its inclusion converts source-count information to an event rate. The co-moving volume element:
\vspace{-0.5mm}
\begin{equation}\label{dvdz}
\frac{\mathrm{d}V}{\mathrm{d}z}= \frac{4\pi c}{H_{0}}\frac{d_\mathrm{\hspace{0.25mm}L}^{\hspace{1.5mm}2}(z)}{(1 +
z)^{\hspace{0.25mm}2}\hspace{0.5mm}h(z)}\,,
\end{equation}

\vspace{-0.5mm}
\noindent describes how the number densities of non-evolving objects locked into Hubble flow are constant with redshift. The quantity $h(z)$, is the normalized Hubble parameter,
\vspace{-0.5mm}
\begin{equation}\label{hz}
h(z)\equiv H(z)/H_0 = \big[\Omega_{\mathrm m} (1+z)^3+ \Omega_{\mathrm \Lambda} \big]^{1/2}\,,
\end{equation}

\noindent where $\Omega_{\mathrm m} + \Omega_{\mathrm \Lambda}=1$ \citep[for further details see][]{Carroll_ARAA_1992}. For a `flat-$\Lambda$' cosmology, we take $\Omega_{\mathrm
m}=0.3$, $\Omega_{\mathrm \Lambda}=0.7$ and
\mbox{$H_{0}=72$ km s$^{-1}$ Mpc$^{-1}$} for the Hubble parameter at the present epoch.

\section{The observation-time relation for peak flux and redshift}

\subsection{The log\,$P$--log\,$T$ relation}
\label{section_logPlogT_theory}

From equation \ref{dN}, the rate of GRBs with a peak photon flux greater than $P$ observed by an instrument with sky coverage $\Omega$ is given by:
\vspace{-0.5mm}
\begin{equation}
\label{eq_integral_peak_flux}
\hspace{-3mm}
\dot{N}(>P)=\frac{\Omega}{4 \pi} \hspace{-4mm}\int \limits_{\hspace{4mm}L_{\mathrm{min}} }^{\hspace{6mm}L_{\mathrm{max}} } \hspace{-4mm} \Phi(L) \mathrm{d}L \hspace{-12mm}\int \limits_{\hspace{1mm}0}^{\hspace{12mm}z_{\mathrm{max}}(L,P)}
\hspace{-1mm}\frac{\mathrm{d}V (z) }{\mathrm{d}z } \frac{ R_{\mathrm{GRB}}(z) }{( 1 + z)}\,\mathrm{d}z \,,
\vspace{0mm}
\end{equation}

\noindent were $z_{\mathrm{max}}(L,P)$ is the maximum redshift from which a burst with luminosity $L$ and peak flux $P$ can be detected.

To introduce an observation-time dependence, $T$, we follow the probability event horizon concept of \citet{coward_PEH_05} and note that as GRBs are independent of each other their observation-times will follow a Poisson distribution in time. Therefore, the temporal separation between events will follow an exponential distribution defined by a mean number of events, $\dot{N}(> P)\,T$. The probability $\epsilon$ for at least one event $> P$ is given by:
\vspace{-0.5mm}
\begin{equation}\label{eq_peh}
\mathcal{P}(n \ge 1;\dot{N}(> P),T)=  1 - e^{\dot{N}(> P) T} = \epsilon \,.
\end{equation}

\noindent For this equation to remain satisfied with increasing observation-time:
\vspace{-0.5mm}
\begin{equation}\label{eq_logPlogT_cos}
\dot{N}(> P) T =  |\mathrm{ln}(1 - \epsilon)| \,.
\end{equation}

\noindent Equating the above equation for $P$ and $T$ we obtain a relation for the evolution of peak flux as a function of observationtime. By setting $\epsilon$ to some arbitrary value, log\,$P$--\,log $T$ curves can be obtained numerically through equations \ref{eq_integral_peak_flux} and \ref{eq_logPlogT_cos}.

Assuming a constant radial distribution of sources, \citet{howell07_APJL} derived the following compact form for a log\,$P$--\,log $T$ relation, which is a $T^{2/3}$ power law for a Euclidean distribution of sources:

\vspace{-0.5mm}
\begin{equation}
\!\!\!P(T)\!\! =T^{2/3}\!\left(\frac{\Delta\Omega \hspace{0.5mm} r_{0}}{3 \sqrt{4 \pi}\hspace{0.5mm}|\mathrm{ln}(1 -
\epsilon)|}\right)^{\hspace{-1mm}2/3}\hspace{-1mm}\left[\hspace{-4mm} \int\limits_{\hspace{5mm}L_{\mathrm{min}} }^{\hspace{6mm}L_{\mathrm{max}}} \hspace{-5mm}\Phi(L)L^{3/2}dL\right]^{2/3}\hspace{-4mm}.
\label{eq_logPlogT_euc}\\
\end{equation}

\begin{figure}
  \includegraphics[scale = 0.55,bbllx = 100pt,bblly =310pt, bburx = 500 pt, bbury = 570 pt,origin=lr]{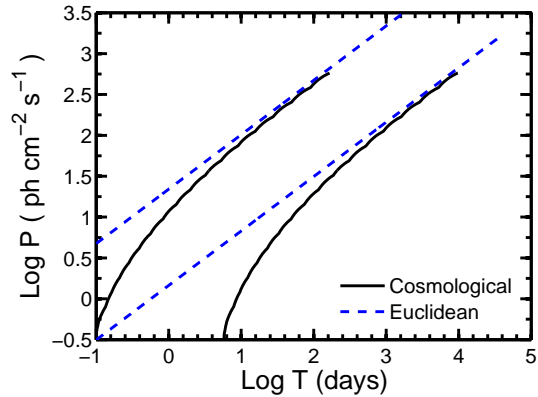}\\
  \caption{Cosmological and Euclidean log\,$P$\,--\,log\,$T$ curves. The two curves converge at high observation-times.}
  \label{logPlogT_curves}
\end{figure}

 Figure \ref{logPlogT_curves} compares both the cosmological (equations \ref{eq_integral_peak_flux} and \ref{eq_logPlogT_cos}) and Euclidean (equation \ref{eq_logPlogT_euc}) curves for arbitrary parameters of source evolution and LF. Following \citet{howell07} we have plotted upper and lower thresholds by setting $\epsilon=(0.95;0.05)$ -- we will refer to these curves as 90\% PEH thresholds. We note that as observation-time increases the Euclidean and Cosmological curves begin to converge. This represents the increased probability of obtaining a cosmological bright event from the bright $-3/2$ power law part of the log $N$--log\,$P$ distribution. The Euclidean curves are convenient in that they can provide initial estimates by fitting to the brightest sample of PEH peak flux data.

\subsection{The log\,$Z$--\,log $T$ relation}
\label{section_logZlogT_theory}

One can extend the arguments of the previous section to derive a log\,$Z$--\,log $T$ relation. From equation \ref{dN} the rate of GRBs observed by an instrument with sky coverage $\Omega$ within a redshift limit $Z_{\rm{L}}$ is given by:

\begin{equation}
\label{eq_integral_z}
\vspace{6mm}
\hspace{-2mm}\dot{N}(< Z_{\rm{L}})=  \hspace{-1mm}\frac{\Omega\, \eta_{z}}{4 \pi} \hspace{-12mm}\int\limits_{\hspace{12mm}L_{\mathrm{min}}(\mathrm{P_{L}},Z_{\mathrm{L}}) }^{\hspace{7mm}L_{\mathrm{max}} } \hspace{-10mm}\Phi(L)\,\mathrm{d}L \hspace{-2mm}
\int\limits_{0}^{\hspace{5mm}Z_{\rm{L}}}
\hspace{1mm}\frac{\mathrm{d}V (z) }{\mathrm{d}z } \frac{ R_{\mathrm{GRB}}(z) }{( 1 + z)}\,\mathrm{d}z \,,
\vspace{-2mm}
\end{equation}

\noindent with $z_{\mathrm{L}}$ obtained by applying the flux limit of the detector, $P_{\mathrm{L}}$, to equation \ref{eq_peakflux}. The quantity $\eta_{z}$ is the efficiency of obtaining a redshift, approximated as the fraction of the total burst sample with measured redshifts.
Using a similar argument as used to determine equation \ref{eq_logPlogT_cos}, one obtains the following relation for the temporal evolution of redshift:
\vspace{-0.5mm}
\begin{equation}\label{eq_logZlogT_cos}
\dot{N}(< Z_{\rm{L}}) T =  |\mathrm{ln}(1 - \epsilon)|. \\
\end{equation}

\noindent This equation can be equated for $T$ and $z$ to set a spatial dependence on GRB populations. Curves of log\,$Z$--\,log $T$ for $\epsilon=(0.95;0.05)$ can be obtained numerically through equations \ref{eq_integral_z} and \ref{eq_logZlogT_cos}.

By using a single log $Z$--log\,$T$ curve as a threshold \citet{coward_PEH_05} showed that SL-GRBs appeared as outliers, supporting the suggestion that they were a sub-population of classical LGRBs. In this paper we have extended their relation, based solely on the cosmological event rate evolution $R_{\mathrm{GRB}}(z)$, to include the effects of detector sensitivity and the luminosity distribution of sources. As shown above, this allows the log $Z$--log\,$T$ relation to be derived seamlessly from a standard integral distribution. Thus, model parameters obtained by fitting to a differential brightness distribution should satisfy the two observation-time relations presented above (equations \ref{eq_logPlogT_cos} and \ref{eq_logZlogT_cos}). This circumvents a difficulty encountered in previous studies which fitted directly to the PEH data to determine a rate -- the resulting rate estimates had low resolution due to the intrinsic scatter of PEH data \citep{howell07,howell:203}.

Before we go on to apply the fits described in sections \ref{section_logPlogT_theory} and \ref{section_logZlogT_theory} to PEH data, we will first determine the parameter values that will be used in the models of \mbox{log\,$(P;Z)$--\,log $T$}. We will do this by fitting our data to the brightness distribution of GRBs.

\section{Fitting to the GRB brightness distribution}
\label{section_logNlogP_results}

In this section we perform $\chi^{2}$ minimisation of the brightness distribution of both \emph{Swift} and BATSE LGRBs to obtain values for the three free parameters $\rho_{0}$, $L_{0}$ and $\alpha$. A goodness of fit for this procedure is obtained by dividing the minimised $\chi^{2}$ value by the number of degrees of freedom, $\chi^{2}/\rm{dof}$. We note that although the \mbox{log\,$(P;Z)$--\,log $T$} relations are derived from an integral distribution, we fit to a differential distribution in which the number of sources are independent at each interval of $P + \mathrm{d}P$\,\footnote{Bright objects will contribute to counts at all values of $P$ in an integral distribution}. From equation \ref{dN} one can define a differential log\,$N$--\,log\,$P$ relation \citep{kommers_2000,pm01,Salvaterra2007,Campisi2010} which is the observed rate of bursts within a peak flux interval ($P_{1}$, $P_{2}$) as:
\vspace{-0.5mm}
\begin{equation}\label{eq_differential_peak_flux}
    \hspace{-1.0mm}
    \dot{N}(P_{1}\hspace{-0.5mm}\leq \hspace{-1.0mm}P\hspace{-1.0mm}< \hspace{-0.5mm}P_{2}) = \hspace{-1mm}\frac{\Omega}{4 \pi}\hspace{-2mm}\int\limits^{\hspace{3mm}\infty}_{\hspace{2mm}0} \frac{\mathrm{d}V (z) }{\mathrm{d}z } \frac{ R_{\mathrm{GRB}} }{( 1 + z)}\mathrm{d}z \hspace{-6.5mm} \int\limits^{\hspace{7mm}L(P_{2},z)}_{\hspace{6mm}L(P_{1},z)} \hspace{-4mm}\Phi(L) \mathrm{d}L\,,
\end{equation}

\noindent with ${L(P_{1,2},z)}$ obtained though equations \ref{eq_rest_frame_luminosity} and \ref{eq_peakflux}.

We bin peak flux data into logarithmical spaced intervals $\Delta P$ and ensure each bin contains at least 5 bursts \citep{PracStatsAst03}. Bursts per bin $\Delta N$ and their uncertainties $\pm \sqrt{\Delta N}$ are converted into burst rates $\Delta R$ by dividing by the live time of the search $\Delta T$ \citep{kommers_2000}.

\subsection{The BATSE log $N$ -- log $P$ distribution}

A number of studies have utilized the \citet{kommers_2000} differential peak flux distribution data \citep[eg.][]{pm01,Salvaterra2007,Campisi2010}. As we require both the discrete peak flux values and their observation-times to produce a $P(T)$ distribution in the next section, we make our own data selection from the BATSE current catalogue\footnote{\url{http://www.batse.msfc.nasa.gov/batse/grb/catalog/current/}}. We note however, the \citet{kommers_2000} sample is a good consistency test for our models and we find good agreement using the best fit parameters of \citep{pm01,Salvaterra2007}.

\begin{figure}
    \includegraphics[scale = 0.55,bbllx = 100pt,bblly =280pt, bburx = 500 pt, bbury = 600 pt,origin=lr]{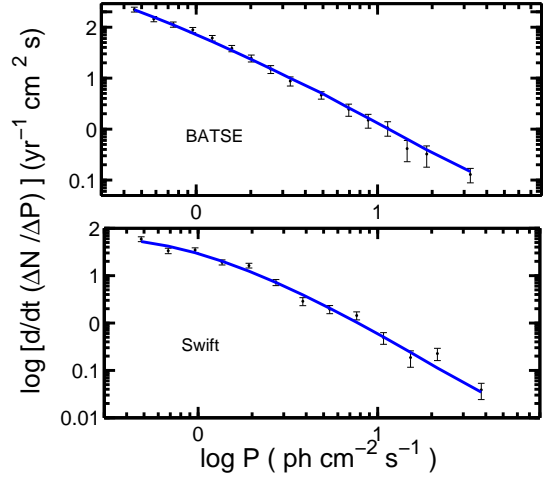}\\
  \caption{The differential log $N$--log $P$ distribution of BATSE (top) and \emph{Swift} (bottom). }
  \label{BATSESwift_logNlogP}
\end{figure}

The BATSE instrument performed a number of runs at different trigger energy ranges and photon count rate thresholds throughout its operation. These runs, in effect, represented different experiments\footnote{For example, during the first 17 months of operation the energy range was set to (50–-300)keV, although the trigger thresholds were varied.}. For consistency, we condition this data by selecting bursts from the (50--300)\,keV band obtained when the trigger thresholds were set to 5.5$\sigma$. Following \citet{guetta05} and \citet{guetta07} we additionally select bursts for which the ratio of count rates, $C$, at the 1024 ms timescale $C_{\mathrm{max}}/C_{\mathrm{min}} \geq  1$ -- here $C_\mathrm{max}$ is the rate in the second brightest detector. We finally select long bursts with $T_{90} > 2$s and with set a threshold of $P_{\mathrm{min}} = 0.4$ corresponding to the value at which the triggering algorithm is almost 100\% efficient \citep{Batse_4_cat}. We find that 596 GRBs meet these criteria.

Table \ref{table_batse_data} summarizes the different data segments and the total number of bursts meeting the above constraints. We show the trigger numbers (rather than the dates) and the total detector run time to form each segment. In our later analysis we will apply the log\,$P$\,--\,log$\,T$ method to both the combined data and to the largest of the individual segments.

%
%
%
%
%
%

\begin{table}
\begin{centering}
\begin{tabular}[scale=1.0]{cccc}
  \hline
Data        &Trigger         & Number of      &Run time \\
            &numbers         & bursts         &(days)\\
  \hline
A                 &105--178     &      11 &  18   \\
B                 &268--1851     &    164 &  437 \\
C                 &1928--3175     &   198 &  724 \\
D                 &3883--3941     &   14  &  41\\
E                 &5403--5519     &   37  &  76\\
F                 &6102--6764     &   143  &  448\\
G                 &7356--7767     &   29  &  228\\

  \hline
\end{tabular}
\caption[]{The BATSE data segments of bursts that triggered in the (50--300)\,keV band when the trigger thresholds were set to 5.5$\sigma$. We show the number of bursts within each segment that passed our selection criteria and the instrument run time associated with each segment.}
\end{centering}
\label{table_batse_data}
\end{table}

Figure \ref{BATSESwift_logNlogP} (top panel) shows the best fitting results to the BATSE differential peak flux distribution using $\Omega=0.67 \pi$ and a live time of $\Delta T=$ 3.19 yr. The peak flux intervals, number of bursts and burst rates data used for the fit are given in Table \ref{table_batse_lognlogp}. We find a best fit of $\rho_{0} =0.12 ^{+0.1}_{-0.01} \mathrm{Gpc}^{-3} \mathrm{yr}^{-1},
L_{*} =3.1^{+364}_{-1.7}\times 10^{51}\, \mathrm{erg} \mathrm{s}^{-1}
\mathrm{and}\,\, \alpha =-2.2 ^{+0.17}_{-32}$. The goodness of fit is given as $\chi^{2}/dof = 6.6/13 \sim 0.51$.



\subsection{The \emph{Swift} log\,$N$\,--\,log\,$P$ distribution}
\label{subsection_swift_logNlogP}

In recent studies \citet{zhang_openQs_2011} and \citet{Bromberg2012} have suggested that the $T_{90}=2\,\mathrm{s}$ division of short and long GRBs based on the BATSE bimodial distribution \citep{Kouveliotou_1993} is a detector dependent categorisation and therefore not appropriate for \emph{Swift} bursts. Other studies have suggested an intermediate duration class of bursts between these two classes \citep{Horvath2010}. For our LGRB sample, we obtain peak flux values from the Swift online catalogue\,\footnote{\url{http://swift.gsfc.nasa.gov/docs/swift/archive/grb_table/}} but rather than employing a $T_{90}$ cut, we use the catagorisations given in the Jochen Greiner online catalogue (JG) of well localized GRBs\,\footnote{\url{http://www.mpe.mpg.de/~jcg/grbgen.html}}. As the burst catagorisations in this catalogue are subject to review through follow up observations we find it a useful resource to obtain our LGRB sample. We find that from the \emph{Swift} peak flux sample of 649 bursts up to burst 120224A, 644 are catagorised in the JG catalogue. We exclude the three SL-GRBs, 060218, 060505 and 100316D (see section \ref{section_logZlogT_SLGRB} for further discussion of these bursts) and 12 bursts catagorised as SGRB-EE (all but one have $T_{90}>2\,\mathrm{s}$) \citep{norris_2011}. We follow \citet{guetta07}, \citet{Wanderman2010} and \citet{Salvaterra2010} and adopt a simplified approach to account for detector sensitivity by applying a peak flux cut at 0.4 ph sec$^{-1}$ cm$^{-2}$ leaving a total sample of 555 LGRBs.

Figure \ref{BATSESwift_logNlogP} (lower panel) shows the best fitting results to the Swift differential peak flux distribution. The peak flux intervals, number of bursts and burst rates data used for the fit is given in Table \ref{table_swift_lognlogp}. We find best fit results of $\rho_{0} =0.09 ^{+0.01}_{-0.01} \mathrm{Gpc}^{-3} \mathrm{yr}^{-1},
L_{*} =2.0^{+0.2}_{-0.02}\times 10^{52}\, \mathrm{erg}\,\mathrm{s}^{-1}
\mathrm{and}\,\, \alpha =-3.8 ^{+0.2}_{-0.6}$ with a goodness of fit  $\chi^{2}/dof = 19.9/10 \sim$ 2. These values are in agreement with the no-evolution model of \citet{Salvaterra2007}.


\section{The log $P$ -- log $T$ distribution of LGRBs}
\label{section_logPlogT_results}

In this section we apply the estimated parameters, $L_{*}$, $\alpha$ and $\rho_{0}$ to the 90\% PEH log $P$ -- log $T$ thresholds given in section \ref{section_logPlogT_theory} and attempt to constrain peak flux PEH data.


As a distant bright burst could produce a similar peak flux as a closer burst of moderate brightness, an intrinsic scatter will be present in $P(T)$ data. This, in combination with the spatial distribution of bursts, means it is difficult to use a 90\% log\,$P$\,--\,log\,$T$ threshold to separate burst populations. The spatial dependence introduced in section \ref{section_logZlogT_theory} will provide a stronger case for untangling burst populations. The goal of this section is two fold: a) to show that the parameters obtained using the log\,$N$\,--\,log\,$P$ fits are compatible and complementary with the log\,$P$\,--\,log\,$T$ relation; b) to demonstrate how data conditioning methods improve the quality and quantity of the PEH data.

\begin{figure}
  \includegraphics[scale = 0.55,bbllx = 100pt,bblly =280pt, bburx = 500 pt, bbury = 600 pt,origin=lr]{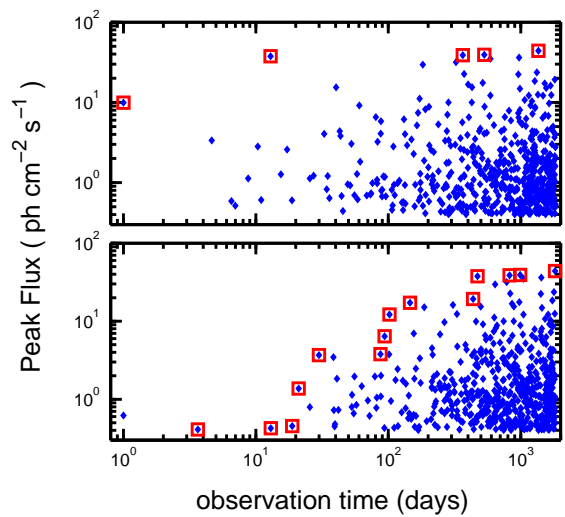}\\
  \caption{The top panel shows the BATSE $P(T)$ sample obtained by chronologically combining the data sets A-F of table \ref{table_batse_data}. The bottom panel shows the same data after applying the FromMAX algorithm to temporally re-ordering the data from the maximum peak flux. The PEH data are indicated in each panel by squares. The top panel illustrates how an energetic event occurring at early observation-time can decreases the PEH sample. The bottom panel shows how some simple processing using FromMAX can circumvent this problem.}
  \label{BATSE_triangle}
\end{figure}

%


\subsection{The BATSE log $P$ -- log $T$ distribution}
\label{BATSE logPlogT data}

The log\,$P$\,--\,log\,$T$ method can be used to fit to either the individual data runs of BATSE shown in Table \ref{table_batse_data} or a combined set -- this allows several consistency tests of the method. We firstly use all the available data to construct a single time series $P(T)$ by chronologically combining the 6 data sets A-F. A temporal dependence is obtained by adding the observation-times of the \emph{j}th data set $T_{j}$ to the maximum of the previous $\mathrm{max}(T_{j-1})$. When joining two consecutive data sets an additional factor of $T=1.5$ days, the mean time between events for BATSE, is also added to the first event of $T_{j}$ to ensure a sufficient delay time is included.


Figure \ref{BATSE_triangle} shows the combined BATSE log $P(T)$ data. The top panel shows the pre-processed sample with PEH data shown by squares. We see that as observation-time increases, so too does the probability of a large event. The plot also illustrates how a bright event can occur near the start of a $P(T)$ time-series (at around 10 days) and thus minimise the PEH sample - we will see later that this effect can also be seen in the individual data sets A-F.


The bottom panel of Figure \ref{BATSE_triangle} shows the same data after applying the fromMAX technique to temporally re-order the data from the brightest event (as described in section \ref{section_data_extraction}). The advantage of the fromMAX method is clearly illustrated -- the PEH sample is increased from 5 events to 14 events.

\begin{figure}
  \includegraphics[scale = 0.55,bbllx = 100pt,bblly =310pt, bburx = 500 pt, bbury = 570 pt,origin=lr]{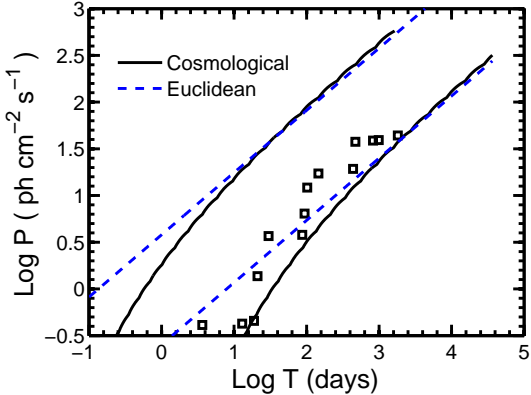}\\
  \caption{The log\,$P$\,--\,log\,$T$ data of Figure \ref{BATSE_triangle} is constrained using cosmological 90\% PEH curves. At high observation-time the data is also constrained by a Euclidean model.}
  \label{batse_logPlogT}
\end{figure}

To test that PEH data set obtained through FromMAX (top panel) is statistically compatible with that obtained from the pre-conditioned data (top panel), we apply a 2D-Kolmogorov-Smirnov test (2DKS) \citep{Peacock83,FF87} and obtain a KS probability of 29\% showing the two samples are statistically consistent.

\begin{figure}
  \includegraphics[scale = 0.55,bbllx = 100pt,bblly =290pt, bburx = 500 pt, bbury = 570 pt,origin=lr]{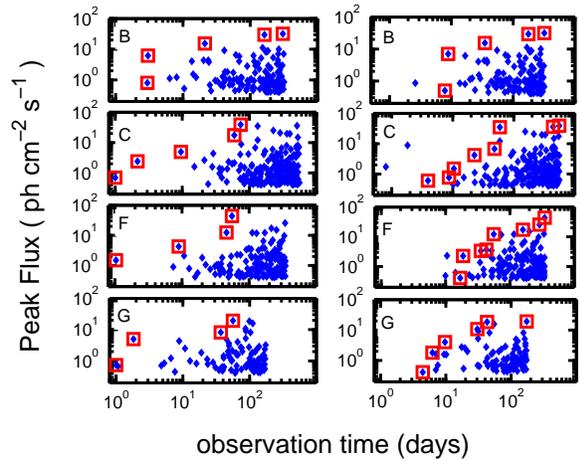}\\
  \caption{As for Figure \ref{BATSE_triangle} but this time showing on the left PEH data from pre-processed $P(T)$ samples B,C,F and G. The FromMAX PEH data from each sample is shown on the right. We see that the sample size is increased in all but the first case. }
  \label{BATSE_triangle_seperate}
\end{figure}

Figure \ref{batse_logPlogT} shows the 90\% log\,$P$\,--\,log\,$T$ thresholds and peak flux PEH data of BATSE. We see that the data is well constrained by the cosmological threshold. In addition to the increased size of the sample, the improvement obtained using the FromMAX method can be further validated through binomial maximum likelyhood (BML) estimates for obtaining data within the 90\% thresholds. We find a BML of 92\% for the data shown in Figure \ref{batse_logPlogT} (1 failure in 14). We find a BML of 60\% for the pre-processed PEH data shown in the top panel of Figure \ref{BATSE_triangle} (2 failures in 5). We note that a bias towards the lower threshold ($\epsilon = 95\%$) is evident - this is a result of the data conditioning rather than an intrinsic bias, as will become apparent when we next apply the method to individual data sets.

Figure \ref{BATSE_triangle_seperate} shows a repeated analysis for the 4 largest individual data segments, B,C,F and G. The left sided panels display the pre-processed data (PEH data is again shown by squares) and the right sided panels show the time series after applying the FromMAX method. We see that the PEH data has increased in 3 of the 4 samples.

\begin{figure}
  \includegraphics[scale = 0.55,bbllx = 100pt,bblly =310pt, bburx = 500 pt, bbury = 570 pt,origin=lr]{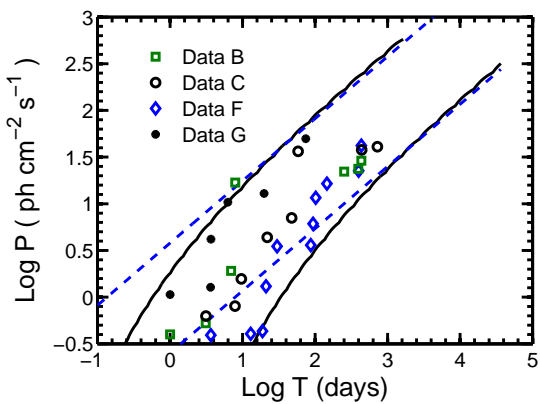}\\
  \caption{As for Figure \ref{batse_logPlogT} but with the PEH data of Figure \ref{BATSE_triangle_seperate} obtained using FromMAX. We see that the PEH data is well constrained by the 90\% log $P$ -- log $T$ thresholds. }
  \label{batse_logPlogT_other}
\end{figure}

Figure \ref{batse_logPlogT_other} shows the 90\% log $P$ -- log $T$ thresholds applied to the 4 sets of PEH data shown in the right hand panels of Figure \ref{BATSE_triangle_seperate}. The plot shows that the PEH data is again well constrained. The bias towards the lower threshold ($\epsilon = 95\%$) evident in Fig. \ref{batse_logPlogT} is only apparent for data set F, suggesting that this is a manifestation of the data conditioning rather than the BATSE sample. The plot does suggest that methods such as data splitting and recombining procedures as used in \citet{howell07} on simulated data could prove useful to increase PEH data sets from long time-series.

Table \ref{table_batse_bml} compares the amount of PEH data and the BML estimations obtained using the 4 sets of pre-processed data with those obtained after applying the FromMAX method. All cases show that the amount of PEH data and/or the BML estimates have improved from the processing. Additionally, the 2DKS probabilities are given, which all indicate good statistical compatibilities between the pre-processed PEH samples and after applying FromMAX.

\begin{table}
  \centering
  \begin{tabular}[scale=1.0]{llcccc}
\hline
\hline
     Data set          &      &  B    &   C   & F    & G \\
\hline
Pre-conditioned & Size & 5     & 5     &4     &4  \\
PEH data        & BML  & 80\%  & 100\% &100\% & 75\% \\
\hline
FromMAX         & Size & 5     & 8     &8      &6 \\
PEH data        & BML  & 100\% & 100\% & 88\% &100\% \\
\hline
2DKS Probability          &      &  0.84  &0.85  &0.68   &0.28 \\
\hline
\hline
\end{tabular}
  \caption{A comparison of PEH data obtained from pre-conditioned data with that obtained using the FromMAX method. The table shows that the amount of data and/or binomial maximum likelyhood (BML) estimates imrove with the FromMAX method.
  A 2D Kolmogorov-Smirnov test on the two PEH samples from each data set shows that the statistical PEH signature is not lost by using the FromMAX method.}
  \label{table_batse_bml}
\end{table}

\subsection{The Swift log\,$P$\,--\,log\,$T$ distribution}

\begin{figure}
  \includegraphics[scale = 0.5,bbllx = 100pt,bblly =290pt, bburx = 500 pt, bbury = 570 pt,origin=lr]{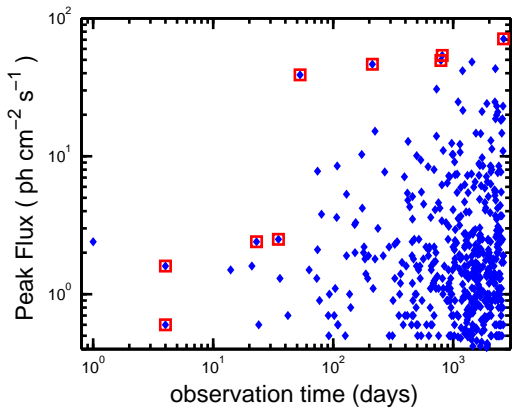}\\
  \caption{The $P(T)$ distribution of \emph{Swift} after applying the FromMAX method.}
  \label{Swift_PT_data}
\end{figure}

Figure \ref{Swift_PT_data} shows the \emph{Swift} FromMAX time-series -- the pre-processed $P(T)$ data was shown previously in Fig. \ref{swift_logPlogT_data}. In comparison to BATSE sample shown in the top panel of Fig. \ref{BATSE_triangle}, one could argue that the Swift data already seems well conditioned to extract PEH data. This is because: a) the largest event in the the pre-processed Swift data occurred near the end of the observation-time; b) no significantly bright event has occurred at early observation-time.

\begin{figure}
  \includegraphics[scale = 0.55,bbllx = 100pt,bblly =300pt, bburx = 500 pt, bbury = 570 pt,origin=lr]{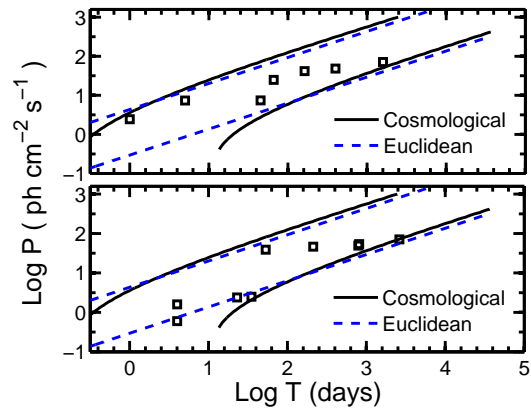}\\
  \caption{The 90\% log\,$P$\,--\,log\,$T$ thresholds for \emph{Swift} are used to constrain PEH data. Top panel - PEH data from the pre-processed time-series. Bottom panel - PEH data using the FromMAX time-series.}
  \label{Swift_logPlogT1}
\end{figure}

Figure \ref{Swift_logPlogT1} shows the 90\% log\,$P$\,--\,log\,$T$ thresholds using both the pre-processed and FromMAX time-series. Both PEH samples recorded BML values of 100\% although the FromMAX method has produced an increase in data. The statistical compatibility between the PEH samples before and after applying FromMAX is confirmed by a value of 0.46.

We have shown in this section that a log\,$P$\,--\,log\,$T$ distribution of GRBs is in agreement with a log\,$N$\,--\,log\,$P$ distribution and that the method can be applied to data sets obtained from different detectors. We have further shown how some processing can improve the quality and quantity of data. In the next section we shall apply the same principles to the redshift domain and will show how the technique can be used to differentiate between different populations of burst.

\section{The \emph{Swift} log\,$Z$\,--\,log\,$T$ distribution}
\label{section_logZlogT_results}

As discussed earlier, the PEH sample of a $Z(T)$ distribution approaches the low-$z$ regime rapidly \citep{coward_PEH_05}, thus the log $Z$--log $T$ method is most sensitive to the closest occurring events. The dependence on only the closest events mean that selection effects such as the `redshift dessert' \citep{coward_GRBDessert_08} can be ignored. Additionally, as the source rate evolution is reasonably well predicted within $z\sim 2$ the choice of SFR model will not change the curves significantly. For the closest events, as redshift determinations would be expected to be more accurate, we include values obtained through all methods: absorbtion, emission and photometrically.

\begin{figure}
  \includegraphics[scale = 0.55,bbllx = 100pt,bblly =270pt, bburx = 500 pt, bbury = 590 pt,origin=lr]{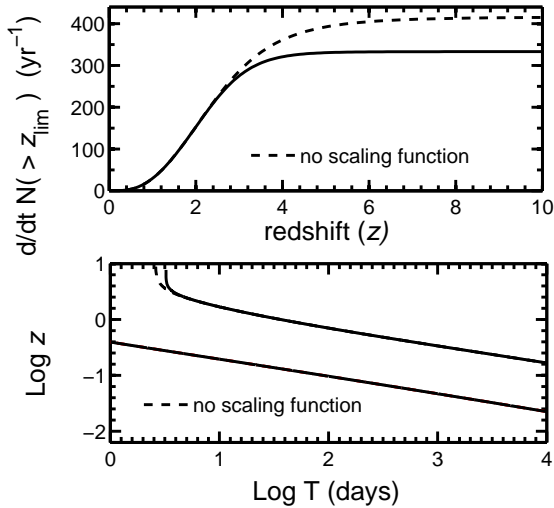}\\
  \caption{The top panel shows the function $\dot{N}(< Z_{\rm{L}})$ with and without the scaling factor discussed in the text. The lower panel shows that the two log\,$Z$\,--\,log\,$T$ curves differ little and are dependent only on the closest events of a source population. }
  \label{logZlogT_sf}
\end{figure}

Figure \ref{logZlogT_sf} illustrates how the high-$z$ variations have minimal influence on the log\,$Z$\,--\,log\,$T$ curves by examining the redshift selection function. The top panel shows the function $\dot{N}(<\,Z_{\rm{L}})$ defined by equation \ref{eq_integral_z} assuming $P_{\mathrm{L}}=0.4$ -- one may recall that log\,$Z$\,--\,log\,$T$ curves are produced by introducing a temporal dependence (equation \ref{eq_logZlogT_cos}). The dashed line shows the same function without the factor $S(L) = \int_{L_{\mathrm{min}}(\mathrm{P_{L}},z) }^{L_{\mathrm{max}} }\!\! \Phi(L) \mathrm{d}L $ which can be referred to as a detector dependent scaling factor \citep[see eq. 1 of][]{coward_GRBDessert_08}. The plot shows that this function has little effect until $z \sim 3$ ($z \sim 2$ for the BATSE instrument) -- as the detection threshold decreases this value of $z$ increases.

The lower panel shows the log $Z$--log $T$ for these two scenarios showing that there is little change in the two curves. Therefore one can make reasonable estimates without an accurate description of the luminosity distribution of the sources. For a log $Z$--log $T$ function, the rate density is the dominant variable. As such, one can discriminate between transient populations of different rate density solely by their spacial dependence.

\begin{figure}
  \includegraphics[scale = 0.55,bbllx = 100pt,bblly =280pt, bburx = 500 pt, bbury = 590 pt,origin=lr]{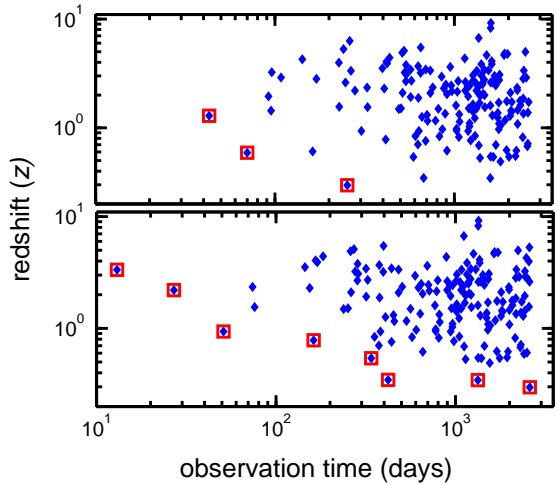}\\
  \caption{The Swift $Z(T)$ distribution with PEH data indicated by squares. The top panel shows the pre-processed sample. The lower panel shows the PEH data has increased after applying the FromMIN procedure.}
  \label{swift_logZlogT_data}
\end{figure}

Figure \ref{swift_logZlogT_data} shows the \emph{Swift} $Z(T)$ distribution with PEH data indicated by squares. The \emph{Swift} $Z(T)$ sample of 173 bursts is obtained by taking bursts from the selection of section \ref{subsection_swift_logNlogP} with certain redshift measurements. The data is given in Table \ref{table_pehdata}. The top panel shows the pre-processed data and the lower panel shows the data after applying the FromMIN procedure described in section \ref{section_data_extraction}. The Figure shows that the PEH data set has been increased by applying FromMIN (the PEH data is indicated by bold in Table \ref{table_pehdata}).

\begin{figure}
  \includegraphics[scale = 0.55,bbllx = 100pt,bblly =310pt, bburx = 500 pt, bbury = 570 pt,origin=lr]{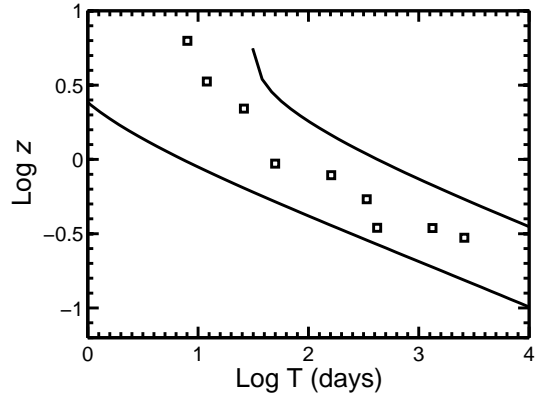}\\
  \caption{The 90\% log\,$Z$\,--\,log\,$T$ thresholds and PEH data for the Swift LGRB sample obtained using the FromMIN method. The data is well constrained using the rate $\rho_{0} = 0.09 ^{+0.01}_{-0.01} \mathrm{Gpc}^{-3} \mathrm{yr}^{-1}$ estimated through a log\,$N$\,--\,log\,$P$ fit. }
  \label{swift_logZlogT_curve}
\end{figure}

Figure \ref{swift_logZlogT_curve} shows that the PEH data is well constrained by the 90\% log $Z$--log $T$ thresholds. To account for the Swift efficiency in obtaining a redshift, $\rho_{0}$ is scaled by a factor of 0.3 to account for the number of bursts with redshifts (172) from our overall sample (576). To test the compatibility of the PEH data obtained using the FromMIN method with that obtained from the pre-processed $Z(T)$ we perform a 2DKS test. A probability of $0.87$ indicates that the two samples are from the same distribution.

We have shown in this section by extracting PEH data from the \emph{Swift} $Z(T)$ sample that the log\,$Z$\,--\,log\,$T$ relation is in good agreement with both the log\,$N$\,--\,log\,$P$ and log\,$P$\,--\,log\,$T$ relations. We have also illustrated that a log\,$Z$\,--\,log\,$T$ distribution relies strongly on the spacial distribution of the sources and is invariant to the detector selection function. In the next section we will demonstrate how the method can differentiate between source populations.

\section{The log\,$Z$\,--\,log\,$T$ relation applied to SL-GRBs}
\subsection{The log\,$Z$\,--\,log\,$T$ relation as a prob of burst populations}

In this section we will demonstrate that the log\,$Z$\,--\,log\,$T$ method can be used to differentiate between different source populations. A key principle we will use is that PEH data represents the rarer, brighter or closer events that are less subject to detector selection bias. Thus, insight into the average properties of a source population can be provided from only a small sample of PEH events.

Bursts from the same distribution, with the same intrinsic rate density, should produce PEH data that is constrained by a log\,$Z$\,--\,log\,$T$ threshold. A separate burst population, occurring at a higher rate, will produce rarer PEH events at earlier observation-times and visa-versa for populations at lower rates. For two mixed $Z(T)$ populations, their observation-time sequence will differ, thus producing outlying events to a 90\% log\,$Z$\,--\,log\,$T$ threshold.

\label{section_logZlogT_SLGRB}
\subsection{The \emph{Swift} sub-luminous GRB sample}

In addition to the two excepted SL-GRBs in the \emph{Swift} sample, GRB 060218 and GRB 100316DA, a number of other candidates have also been discussed. Based on their low luminosities and subsequent Poisson detection probabilities, \citet{Wanderman2010} suggested three additional bursts could contribute to the SLGRB population: GRB 050724, GRB 051109 \citep[see also][]{Bromberg2012} and GRB 060505. Of these three bursts we include GRB 060505 in our sample. Although GRB 050724 was a long duration burst ($T_{90} \sim 96$\,s), further analysis has suggested that it is of the SGRB category \citep{Campana2006,Grupe2006,Malesani2007}-- most likely a SGRB-EE \citep{norris_2010, norris_2011}. GRB 051109 is also omitted due to uncertainty in the redshift estimation\footnote{A putative host galaxy at $z=0.08$ was reported in GCN 5387 by D. A. Perley et. al.} (in accordance with our selection criteria of section \ref{swift_logZlogT_data}).

\begin{table}
  \begin{centering}
  \begin{tabular}{|l|c|c|c|c|c|c|}
  \hline
    \hline
  \hspace{-1.5mm}GRB & $F_{\mathrm{peak}}$ & $T_{90}$ & $T$ & $L_{\rm{iso}}$ & $\alpha$ & z \\
  & ph \hspace{-1.5mm} s$^{-1}$cm$^{2}$ & s & days & erg\,s$^{-1}$ &  & \\
    \hline
      \hline
  \hspace{-1.5mm}060218 & 0.25 & 2100&427 &  $1.4\times 10^{47}$ & -2.3 & 0.03 \\
  \hspace{-1.5mm}060505 & 2.65 & 4 &504 & $9.1\times 10^{48}$  & -1.3 & 0.089 \\
  \hspace{-1.5mm}100316D & 0.1 & 1300 &1915& $4.8\times 10^{47}$ & -2.3 & 0.059 \\
  \hline
\end{tabular}
\end{centering}
  \caption{Data for the three \emph{Swift} sub-luminous bursts. The isotropic rest frame (1-10000 keV) luminosities
  are obtained from the peak luminosity values (ph s$^{-1}$) using a simple power law fit with values of $\alpha$ taken from the \emph{Swift} online catalogue.}
  \label{table_SLGRB}
\end{table}

Despite GRB 060505 being relatively nearby and well observed, intense photometric and spectroscopic searches found no evidence of an associated supernova despite the fact an event $\sim$100 times fainter than SN 1998bw would have been detected \citep{McBreen2008ApJ, Xu2009ApJ}. Dust obscuration was excluded for this burst and \citet{Fynbo20061047} and \citet{Ofek2007ApJ} argued that the simplest explanation was that GRB 060505 was simply the closest observed SGRB. Other studies however found the properties of the host galaxy consistent with that of the LGRB class \citep{Ofek2007ApJ, Thoene2007, Thoene2008ApJ}. Furthermore, \citet{McBreen2008ApJ} showed the spectral lag was consistent with that of a LGRB\footnote{ LGRBs support the lag–luminosity relation of \citet{Norris2000ApJ} while SGRBs have zero lag.} and an extensive study of the afterglow performed by \citet{Xu2009ApJ} found afterglow parameters within the range for other LGRBs. In view of its luminosity, we include this burst from our analysis of SL-GRBs in the next section. However, to further address the classification of this burst, we will investigate its compatibility to the Swift SGRB population in the final section. Table \ref{table_SLGRB} shows the main parameters for these three SL-GRBs which will be used in the analysis of this section.

\subsection{The log $Z$ -- log $T$ analysis of \emph{Swift} SL-GRBs}

\begin{figure}
  \includegraphics[scale = 0.55,bbllx = 100pt,bblly =300pt, bburx = 500 pt, bbury = 570 pt,origin=lr]{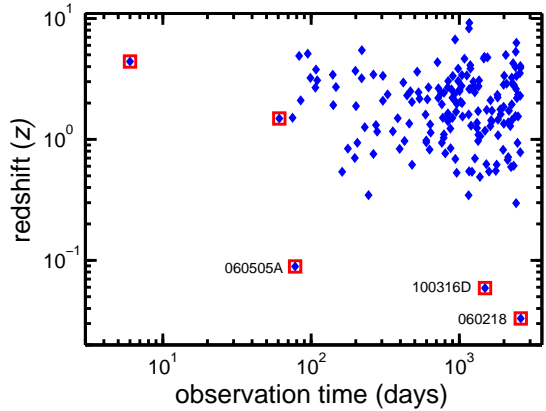}\\
  \caption{The $Z(T)$ time series for Swift LGRBs including three \emph{Swift} SL-GRBs.}
  \label{swift_logZlogT_PTdatawithSLGRB}
\end{figure}

Figure \ref{swift_logZlogT_PTdatawithSLGRB} shows the $P(T)$ time-series for the LGRB sample obtained using FromMIN and including the three SL-GRBs 060218, 060505 and 100316D. The PEH data are indicated by squares -- the SL-GRB sample are clearly shown as significantly close events. On inspection, GRB 060505 stands out as a particularly rare event due to its occurrence at a relatively early observation-time.

\begin{figure}
  \includegraphics[scale = 0.55,bbllx = 100pt,bblly =300pt, bburx = 500 pt, bbury = 570 pt,origin=lr]{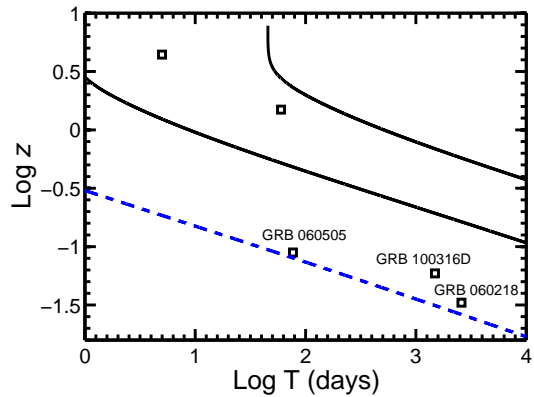}\\
  \caption{The log $Z$--log $T$ distribution for \emph{Swift} LGRBs including three SL-GRBs which are labeled. The dashed line represents a 0.02\% probability for obtaining a LGRB within redshift $z$ at an observation-time $T$. }
  \label{swift_logZlogT_withSLGRB}
\end{figure}

Figure \ref{swift_logZlogT_withSLGRB} shows the 90\% log\,$Z$\,--\,log\,$T$ thresholds based on the LGRB parameters used in section \ref{section_logZlogT_results} along with the PEH data. This plot clearly shows that the three SL-GRBs are outliers to the distribution suggesting they are from a population occurring at a higher rate. Based on the 90\% thresholds of Fig. \ref{swift_logZlogT_curve}, assuming that the SL-GRBs are from the low-probability tail of the LGRB distribution, one would expect to wait 87 (1800) years to observe a burst as close as GRB 060505 (GRB 060218). The probability of observing these bursts at time $T$ can be estimated by extending the PEH threshold upper limit through $\epsilon$ - this new limit is shown in Fig. \ref{swift_logZlogT_withSLGRB} by the dashed curve. We find a probability value of $\epsilon = 0.00015$ is required to constrain the SL-GRB bursts at the LGRB rate. This value is consistent with that estimated by \citet{coward_LLGRB_05} for the probability of observing the low-$z$ burst GRB 980425.

If the SL-GRB sample are from a distinct population one should be able to constrain them using seperate 90\% log\,$Z$\,--\,log\,$T$ thresholds set at a higher rate value $\rho_{0,\mathrm{SL}}$ and a LF representative of the SL-GRB sample. For the LF we consider the form used for the LGRB sample but with $L_{*} = 10^{49}$ erg s$^{-1}$ which is around three orders of magnitude lower than that derived in section \ref{section_logNlogP_results}. As described in the previous section, consideration of $S(L)$ is not essential but is included for completeness. To estimate $\rho_{0,\mathrm{SL}}$ for the \emph{Swift} SL-GRB sample we follow \citet{GuettaDellaValle_2007,CowardHowellPiran_2012} and calculate

\begin{equation}\label{eq_vmax_slgrb}
    \rho_{0,SL} = \sum_{i}^{3} \frac{1}{V_{\rm{max}}} \frac{1}{T} \frac{1}{\Omega} \frac{1}{\eta_{z}}
\end{equation}

\noindent Here $V_{\mathrm{max}}$ is the maximum volume each burst could be detected, $T$ is the maximum observation-time for the sample, $\Omega$ is the sky coverage and we assume $\eta_{z}$ =0.3 as for the LGRB sample. To determine $V_{\mathrm{max}}$
we use equation \ref{eq_rest_frame_luminosity} and assume a flux limit of 0.15 ph sec$^{-1}$ -- this  corresponds with the detection threshold of 95\% of the \emph{Swift} LGRB sample. We obtain $\rho_{0,\mathrm{SL}} =147^{+180}_{-92} \mathrm{Gpc}^{-3}\mathrm{yr}^{-1}$ where the errors are the 90\% Poisson confidence limits \citep{Gehrels_1986}. This rate, based solely on the \emph{Swift} sample, is in agreement with other studies of the SL-GRB population \citep[e.g.][]{coward_LLGRB_05,sodoburg_06_LLGRBRate_06,GuettaDellaValle_2007,Pian_LLGRBs_06,Liang_07,Virgilii_LLGRBs_08,howell_SLGRB_2010}

\begin{figure}
  \includegraphics[scale = 0.55,bbllx = 100pt,bblly =300pt, bburx = 500 pt, bbury = 570 pt,origin=lr]{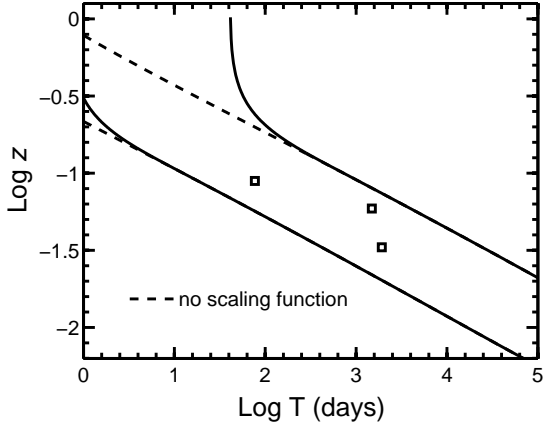}\\
  \caption{The log $Z$--log $T$  distribution for \emph{Swift} SL-GRBs.}
  \label{swift_SLGRB_logZlogT_curve}
\end{figure}

Figure \ref{swift_SLGRB_logZlogT_curve} shows the 90\% log\,$Z$\,--\,log\,$T$ thresholds corresponding to $\rho_{0,\mathrm{SL}}$. To show that the log\,$Z$\,--\,log\,$T$ method allows constraints to be made solely on $\rho_{0,\mathrm{SL}}$, the dashed lines show the curves obtained without the selection function $S(L)$. The SL-GRB PEH data is well constrained demonstrating good agreement with the rate estimates and supporting the hypothesis of a distinct SL-GRB population of bursts. Additionally, the figure supports the hypothesis that GRB 060505 is a member of the SL-LGRB category.

\subsection{Testing the connection between SL-GRBs and XRFs}

We note that the two sub-luminous bursts GRB 060218 and GRB 100316D are often categorised as XRFs. Although XRFs are generally understood to represent the fainter and softer part of the GRB distribution \citep{Zhang_2007ChJAA,Sakamoto_2008ApJ} the relation between XRFs and SL-GRBs is not clear. Looking at \emph{Swift} data the durations of both GRB 060218 (2100\,s) and GRB 060218 (1300\,s) are both much higher than those of other XRFs in the catalogue\footnote{Of the 11 XRFs in the Swift data up to January 2012 the longest $T_{90}$ is 103s; 6 have durations below 10s.} suggesting that a different mechanism separates the two categories. It is therefore interesting to extend the log\,$Z$\,--\,log\,$T$ analysis of the previous section to investigate the connection between SL-GRBs and XRFs using $Z(T)$ data. We find that including two XRFs with secure redshifts\footnote{As in section \ref{subsection_swift_logNlogP} we again base our catatgorisations and redshifts on the Jochen Greiner online catalogue.}: XRF 050416A (z=0.65) and XRF 050824 (z=0.83); does not change the result given in Figure \ref{swift_logZlogT_withSLGRB}. Our analysis using the \emph{Swift} sample with secure redshifts therefore suggests that XRFs and SL-GRBs are from a different population.


\subsection{A log\,$Z$\,--\,log\,$T$ analysis of GRB 060505}

Although the last section shows that based on a log\,$Z$\,--\,log\,$T$ analysis, GRB 060505 is not of the LGRB category, the lack of an accompanying SN to stringent limits still suggests it could be of the SGRB class. A additional test is therefore to include this event in the $Z(T)$ distribution of \emph{Swift} SGRBs and repeat the previous analysis.

To calculate the observed rate of SGRBs, $\rho_{0,S}$, we
add GRB 060505 to the SGRB sample of \citet{CowardHowellPiran_2012} and extend equation \ref{eq_vmax_slgrb} through the factor $R_{B/S}=6.7$ to account for the reduced sensitivity of \emph{Swift} for detecting SGRBs in comparison to BATSE. We omit an anomalous burst GRB 080905A from our sample -- a repeat analysis at the end of this section will consider this burst. Setting $\eta_{z}=9/41$ we obtain $\rho_{0,\mathrm{S}} = 8.57^{+24.7}_{-7.7} \mathrm{Gpc}^{-3}\mathrm{yr}^{-1}$.

\begin{table}
  \centering
  \begin{tabular}{ccc}
\hline
\hline
\hline
GRB  &     Peak Flux  & redshift  \\
\hline
\hline
101219A  &   4.1  &   0.718  \\
100206A  &   1.4  &   0.4068  \\
100117A  &   2.9  &   0.92  \\
090510   &   26.3  &   0.903  \\
070724A  &   2.03  &   0.457  \\
061217   &   2.36  &   0.827  \\
051221A  &   12  &   0.5465  \\
050509B  &   3.71  &   0.226  \\
\hline
080905A$^{\dag}$  &   6.03  &   0.1218  \\
060505$^{*}$   &   2.65  &   0.089  \\
\hline
\end{tabular}
  \caption{The \emph{Swift} SGRB data sample. The lower two entries are investigated as members of the SGRB population: $^{*}$\,possible SL-GRB but no SN observed to deep limits; $^{\dag}$\,low-$z$ outlier to the Yonetoku relation with high energy properties typical of a SGRB (see section 10.4).}
  \label{table_swift_lognlogp}
\end{table}

\begin{figure}
  \includegraphics[scale = 0.55,bbllx = 100pt,bblly =300pt, bburx = 500 pt, bbury = 570 pt,origin=lr]{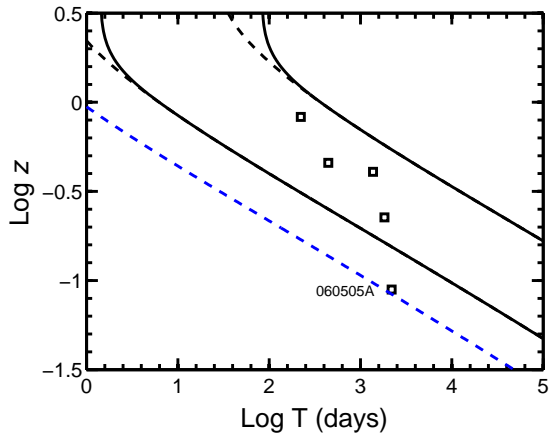}\\
  \caption{The log\,$Z$\,--\,log\,$T$ distribution for Swift SGRBs showing that GRB 060505 is an outlier to this distribution. The dashed curve indicates a $>99\%$ probability that a SGRB would not have occurred at $z=0.09$.}
  \label{swift_logZlogT_curve1}
\end{figure}

Figure \ref{swift_logZlogT_curve1} shows the 90\% log\,$Z$\,--\,log\,$T$ thresholds including GRB 060505, which is shown as a clear outlier. We find that a value of $\epsilon=0.007$ is required to constrain this burst indicating a $>99\%$ probability that a SGRB would not have occurred at $z=0.09$. Our analysis therefore suggests that GRB 060505 belongs in the SL-GRB sample of bursts.

The previous analysis excluded the burst GRB 080905A. This spectrally hard, short duration burst ($T_{90}=1\,$s) showed negligible spectral lag, no evidence of extended emission or of an associated SN to deep limits -- all properties of a SGRB resulting from a compact object merger. It was spectroscopically associated with a galaxy at redshift $z=0.1218$ making it the closest possible SGRB \citep{Rowlinson2010MNRAS,Kann2011ApJ}.

\begin{figure}
  \includegraphics[scale = 0.55,bbllx = 100pt,bblly =300pt, bburx = 500 pt, bbury = 570 pt,origin=lr]{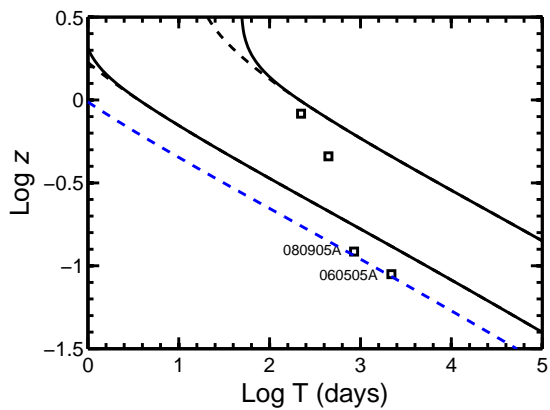}\\
  \caption{As for Fig. \ref{swift_logZlogT_curve1} but including the burst GRB 080905A. We see that this burst, as well as GRB 060505 are outliers. The dashed curve indicates a $\sim$ 1\% probability that GRB 080905A could have occurred at $z = 0.12$.}
  \label{swift_logZlogT_curve2}
\end{figure}

Figure \ref{swift_logZlogT_curve2} shows the result of a repeated analysis of GRB 060505 including GRB 080905A with 90\% log\,$Z$\,--\,log\,$T$ thresholds corresponding to a recalculated rate of $\rho_{0,\mathrm{S}} = 13.5^{+39.1}_{-12.2} \mathrm{Gpc}^{-3}\mathrm{yr}^{-1}$. The plot shows that as well as GRB 060505, GRB 080905A is also an outlier to the 90\% thresholds. This finding supports an analysis by \citet{Gruber2012} which found GRB 080905A to be a clear outlier to the Yonetoku, E$_{\mathrm{p},\mathrm{rest}}$--L$_{\mathrm{p}}$ relation \citep{Yonetoku2004ApJ}; but see also \citet{Borgonovo_2006ApJ}. They found that GRB 080905A would require a redshift $z \sim 0.9$ to be consistent with the Yonetoku relation, suggesting that the host, a large almost face-on spiral galaxy, could be a foreground galaxy. \citet{Rowlinson2010MNRAS} noted that the offset of the afterglow was large (18.5 kpc) but comparable to other SGRB locations, especially considering the relative size of the host.

Our analysis suggests a $\sim$ 1\% probability (indicated by the dashed curve) that GRB 080905A could have occurred at $z = 0.12$. Furthermore, in agreement with \citet{Gruber2012}, we find that given a redshift of $z \sim 0.9$, this burst would be constrained by the log\,$Z$\,--\,log\,$T$ relation.

Given that the observed prompt properties of this burst are highly consistent with a SGRB and the redshift is reasonable secure, the above analysis is interesting and implies that this burst may deserve additional attention. It certainly further highlights the ability of the log\,$Z$\,--\,log\,$T$ method to select out bursts with anomalous properties from a given sample.
\vspace{-5mm}
\section{Conclusions}
In this study we have shown how the redshift and peak-flux distributions of GRBs have an observation-time dependence and have derived two new relations from the standard integral distributions - the log\,$P$\,--\,log\,$T$ and the log\,$Z$\,--\,log\,$T$. We have shown how a PEH data set can be extracted from the redshift and peak-flux time series and constrained by 90\% thresholds using the above two relations. We have demonstrated that the FromMIN and FromMAX methods improve the PEH data sample and circumvent a bias introduced by the occurrence early rare event.\\
\indent By applying the log\,$P$\,--\,log\,$T$ method to both \emph{Swift} and BATSE data we showed that the relation is in agreement with parameters obtained through a log\,$N$\,--\,log\,$P$ fit. We then showed that the same parameters could be used to constrain PEH data using a log\,$Z$\,--\,log\,$T$ relation. By including the Swift SL-GRB sample of bursts we showed that this sample were outliers to the LGRB sample and could be constrained by a 90\% log\,$Z$\,--\,log\,$T$ threshold at a higher rate. We suggest that this is further evidence that SL-GRBs are a discrete population of bursts.\\
\indent We have shown that the methods presented here are strongly dependent on the event rate density of the sources and a good description of the scaling function $S(L)$ is not essential. Additionally, in the redshift regime where the method is dependent on only the closest events, high-$z$ biases do not effect the analysis. Used in the peak flux regime, the method depends on only the brightest of bursts which are less likely to be missed because of detection thresholds. We suggest that an observation-time analysis is a useful complement to other methods that use the whole distribution and can be a good indicator of selection bias.\\
\indent The log\,$P$\,--\,log\,$T$ method, in comparison with a log\,$Z$\,--\,log\,$T$ technique can not be used to discriminate between populations occurring at different rates. A close burst could arise from the low luminosity tail of the LF and equivalently a distant burst would be intrinsically bright. Therefore a log\,$Z$\,--\,log\,$T$ relation which has only a spacial dependence is a stronger indicator. As an intrinsic luminosity is determined from both the redshift and peak flux, a log\,$L$\,--\,log\,$T$ relation, which includes a measure of both the energy and spatial distribution could prove useful. Such a relation may be useful to untangle populations such as SGRBs and SGRB-EEs which have similar redshift distributions.\\
\indent We analysed the burst GRB 060505 using the log\,$Z$\,--\,log\,$T$ relation to determine if it could be of the SL-GRB category. Including this burst with the other two \emph{Swift} SL-GRBs, GRB 060218 and GRB 100316D, we found that all three bursts were constrained at rate $\rho_{0,\mathrm{SL}} = 147 \mathrm{Gpc}^{-3}\mathrm{yr}^{-1}$ estimated independently by a $V_{\mathrm{MAX}}$ analysis. Therefore, in addition to its low-luminosity and relative close proximity a log\,$Z$\,--\,log\,$T$ analysis supports the hypothesis that this burst is a SL-GRB. In contrast to other SL-GRBs, despite intense photometric and spectroscopic searches, this burst showed no evidence of an associated supernova. To tests its compatibility with the SGRB population, we performed a further log\,$Z$\,--\,log\,$T$ analysis and found a greater than 99\% probability that a SGRB would not have occurred as close as this burst.\\
\indent If the progenitor of GRB 060505 was not a compact object merger, upper limits on the ejected $^{56}$Ni mass of $M(^{56}$Ni)$\sim 10^{-3}M_{\odot}$ \citep{Tominaga2007ApJ} support the possibility that this burst may have been accompanied by a low energy SN undergoing significant fallback \citep{Woosley1995ApJS,Fryer_2006ApJ}.\\
\indent An additional analysis of the burst GRB 080905A shows how the log\,$Z$\,--\,log\,$T$ relation can select out bursts with anomalous properties from a given sample. Using the method in a bootstrapping type scheme may be a useful consistency test for both the completeness of sample selections and some of the spectral energy relations related to the observed quantities of GRBs \citep[e.g.][]{Amati2002AandA,Yonetoku2004ApJ,Ghirlanda2005MNRAS}.

\section*{Acknowledgments}
E. J. Howell acknowledges support from a UWA Research Fellowship. D.M. Coward is supported by an Australian Research Council Future Fellowship. The authors thank the referee for a thorough reading of the manuscript and for providing a number of insightful suggestions.

\appendix
\section{The differential peak flux distribution data of BATSE and \emph{Swift} LGRBs}
\label{appendix_batse_data}

\begin{table}
  \begin{tabular}[]{cccc}
\hline
\hline
      P1  & P2  &     $\dot{N}$   &   $\Delta \dot{N}/ \Delta P$  \\
      $ \mathrm{ph}\, \mathrm{sec}^{-1}\, \mathrm{cm}^{-2}$   & $ \mathrm{ph}\, \mathrm{sec}^{-1}\, \mathrm{cm}^{-2}$ &  $\mathrm{yr}^{-1}$     &  $\mathrm{yr}^{-1}\, \mathrm{ph}\, \mathrm{sec}\, \mathrm{cm}^{2}$ \\
\hline
\hline
0.4  &  0.51 & 25 &  222\\
0.51  &  0.66 & 22 &  145\\
0.66  &  0.84 & 20 &  113\\
0.84  &  1.1 & 17 &  88.2\\
1.1  &  1.4 & 15 &  61\\
1.4  &  1.8 & 13 &  38.4\\
1.8  &  2.3 & 11 &  24.6\\
2.3  &  2.9 & 8.8 &  15\\
2.9  &  3.7 & 7.2 &  8.78\\
3.7  &  6.1 & 11 &  4.64\\
6.1  &  7.8 & 3.8 &  2.44\\
7.8  &  10 & 3 &  1.5\\
10  &  13 & 2.4 &  1.06\\
13  &  16 & 1.9 &  0.415\\
16  &  21 & 1.5 &  0.324\\
21  &  44 & 2.7 &  0.129\\
\hline
\end{tabular}
  \caption{The data used in fitting the differential peak flux distribution of the BATSE long GRB sample}
  \label{table_batse_lognlogp}
\end{table}


\begin{table}
  \centering
  \begin{tabular}{cccccc}
\hline
\hline
      P1  & P2  &     $\dot{N}$   &   $\Delta \dot{N}/ \Delta P$  \\
      $ \mathrm{ph}\, \mathrm{sec}^{-1}\, \mathrm{cm}^{-2}$   & $ \mathrm{ph}\, \mathrm{sec}^{-1}\, \mathrm{cm}^{-2}$ &  $\mathrm{yr}^{-1}$     &  $\mathrm{yr}^{-1}\, \mathrm{ph}\, \mathrm{sec}\, \mathrm{cm}^{2}$ \\
\hline
\hline
0.4  &  0.56 & 9.7 &  59\\
0.56  &  0.8 & 7.8 &  33.4\\
0.8  &  1.1 & 11 &  34.6\\
1.1  &  1.6 & 8.9 &  19.1\\
1.6  &  2.2 & 11 &  16.3\\
2.2  &  3.2 & 6.7 &  7.2\\
3.2  &  4.5 & 3.8 &  2.87\\
4.5  &  6.3 & 3.6 &  1.95\\
6.3  &  8.9 & 3.8 &  1.44\\
8.9  &  13 & 1.8 &  0.49\\
13  &  18 & 0.97 &  0.187\\
18  &  25 & 1.7 &  0.227\\
25  &  50 & 0.97 &  0.0388\\
\hline
\end{tabular}
  \caption{The data used in fitting the differential peak flux distribution of the \emph{Swift} long GRB sample}
  \label{table_swift_lognlogp}
\end{table}


\section{PEH redshift data}
\label{appendix_swift_PEH_redshift_data}

\begin{table*}
 \centering
 \begin{minipage}{140mm}
  \caption{The Swift redshift-observation-time data $Z(T)$ used in the analysis. The data is taken from December 19 2004 and re-ordered as a time-series using the \emph{FromMIN} method. The PEH data is shown as bold.}
  \begin{tabular}{llll|llll|llll}
 GRB  & $T_{\rm{obs}}$  &   redshift  && GRB  & $T_{\rm{obs}}$  &   redshift  & & GRB  & $T_{\rm{obs}}$  &   redshift  &\\
 \hline
\textbf{050904}   &   8 &  6.29 &    &  071117   &   811 &  1.331 &    &  091018   &   1512 &  0.971  & \\
\textbf{050908}   &   12 &  3.344 &    &  071122   &   816 &  1.14 &    &  091020   &   1514 &  1.71  & \\
\textbf{050922C}  &   26 &  2.198 &    &  080129   &   888 &  4.349 &    &  091024   &   1518 &  1.092  & \\
\textbf{051016B}  &   50 &  0.9364 &    &  080207   &   896 &  2.2 &    &  091029   &   1523 &  2.752  & \\
051109A  &   73 &  2.346 &    &  080210   &   899 &  2.641 &    &  091109A  &   1533 &  3.076  & \\
051111   &   75 &  1.55 &    &  080310   &   929 &  2.42 &    &  091127   &   1551 &  0.49  & \\
060115   &   144 &  3.53 &    &  080319B  &   938 &  0.937 &    &  091208B  &   1562 &  1.063  & \\
060124   &   153 &  2.296 &    &  080319C  &   938 &  1.95 &    &  100219A  &   1638 &  4.6667  & \\
\textbf{060202}    &   161 &  0.783 &    &  080330   &   949 &  1.51 &    &  100302A  &   1651 &  4.813  & \\
060206   &   165 &  4.048 &    &  080411   &   960 &  1.03 &    &  100316B  &   1665 &  1.18  & \\
060210   &   169 &  3.91 &    &  080413A  &   962 &  2.433 &    &  100418A  &   1697 &  0.6235  & \\
060223A  &   182 &  4.41 &    &  080413B  &   962 &  1.1 &    &  100425A  &   1704 &  1.755  & \\
060418   &   237 &  1.489 &    &  080430   &   979 &  0.767 &    &  100513A  &   1722 &  4.772  & \\
060502A  &   251 &  1.51 &    &  080520   &   999 &  1.545 &    &  100621A  &   1760 &  0.542  & \\
060510B  &   259 &  4.9 &    &  080603B  &   1012 &  2.69 &    &  100724A  &   1793 &  1.288  & \\
060512   &   261 &  2.1 &    &  080604   &   1013 &  1.416 &    &  100728B  &   1797 &  2.106  & \\
060522   &   271 &  5.11 &    &  080605   &   1014 &  1.6398 &    &  100814A  &   1813 &  1.44  & \\
060526   &   275 &  3.221 &    &  080607   &   1016 &  3.036 &    &  100901A  &   1830 &  1.408  & \\
060604   &   283 &  2.68 &    &  080707   &   1046 &  1.23 &    &  100906A  &   1835 &  1.727  & \\
060605   &   284 &  3.78 &    &  080710   &   1049 &  0.845 &    &  101219B  &   1938 &  0.55  & \\
060607A  &   286 &  3.082 &    &  080721   &   1060 &  2.591 &    &  110106B  &   1960 &  0.618  & \\
060707   &   316 &  3.425 &    &  080804   &   1073 &  2.2045 &    &  110128A  &   1982 &  2.339  & \\
060708   &   317 &  1.92 &    &  080805   &   1074 &  1.505 &    &  110205A  &   1989 &  2.22  & \\
060714   &   323 &  2.711 &    &  080810   &   1079 &  3.35 &    &  110213A  &   1997 &  1.46  & \\
\textbf{060729}   &   338 &  0.54 &    &  080905B  &   1104 &  2.374 &    &  110213B  &   1997 &  1.083  & \\
060814   &   353 &  0.84 &    &  080906   &   1105 &  2.1 &    &  110422A  &   2066 &  1.77  & \\
060904B  &   373 &  0.703 &    &  080913   &   1112 &  6.695 &    &  110503A  &   2077 &  1.613  & \\
060906   &   375 &  3.686 &    &  080916A  &   1115 &  0.689 &    &  110715A  &   2149 &  0.82  & \\
060908   &   377 &  1.8836 &    &  080928   &   1127 &  1.692 &    &  110731A  &   2165 &  2.83  & \\
060912A  &   381 &  0.937 &    &  081007   &   1136 &  0.5295 &    &  110801A  &   2165 &  1.858  & \\
060926   &   395 &  3.2 &    &  081008   &   1137 &  1.9685 &    &  110808A  &   2172 &  1.348  & \\
060927   &   396 &  5.47 &    &  081028A  &   1157 &  3.038 &    &  110818A  &   2182 &  3.36  & \\
061007   &   406 &  1.261 &    &  081029   &   1158 &  3.8479 &    &  111008A  &   2232 &  4.9898  & \\
\textbf{061021}   &   420 &  0.3463 &    &  081118   &   1177 &  2.58 &    &  111107A  &   2261 &  2.893  & \\
061110A  &   439 &  0.758 &    &  081121   &   1180 &  2.512 &    &  111209A  &   2293 &  0.677  & \\
061110B  &   439 &  3.44 &    &  081203A  &   1192 &  2.05 &    &  111228A  &   2312 &  0.714  & \\
061121   &   450 &  1.314 &    &  081222   &   1211 &  2.77 &    &  111229A  &   2313 &  1.3805  & \\
061126   &   455 &  1.1588 &    &  081228   &   1217 &  3.4 &    &  120119A  &   2338 &  1.728  & \\
061222A  &   481 &  2.088 &    &  081230   &   1219 &  2 &    &  050126   &   2381 &  1.29  & \\
061222B  &   481 &  3.355 &    &  090102   &   1226 &  1.547 &    &  050223   &   2408 &  0.5915  & \\
070110   &   504 &  2.352 &    &  090205   &   1259 &  4.6497 &    &  050315   &   2430 &  1.949  & \\
070208   &   532 &  1.165 &    &  090313   &   1297 &  3.375 &    &  050318   &   2433 &  1.44  & \\
070306   &   560 &  1.4959 &    &  \textbf{090417B}  &   1331 &  0.345 &    &  050319   &   2434 &  3.24  & \\
070318   &   572 &  0.836 &    &  090418A  &   1332 &  1.608 &    &  050401   &   2446 &  2.9  & \\
070411   &   595 &  2.954 &    &  090423   &   1337 &  8.26 &    &  050505   &   2480 &  4.27  & \\
070419A  &   603 &  0.97 &    &  090424   &   1338 &  0.544 &    &  050525A  &   2500 &  0.606  & \\
070506   &   620 &  2.31 &    &  090426   &   1340 &  2.609 &    &  050603   &   2508 &  2.821  & \\
070521   &   635 &  1.35 &    &  090429B  &   1343 &  9.2 &    &  050730   &   2565 &  3.967  & \\
070529   &   643 &  2.4996 &    &  090516A  &   1360 &  4.109 &    &  050801   &   2566 &  1.56  & \\
070611   &   655 &  2.04 &    &  090519   &   1363 &  3.85 &    &  050814   &   2579 &  5.3  & \\
070612A  &   656 &  0.617 &    &  090529   &   1373 &  2.625 &    &  050820A  &   2585 &  2.612  & \\
070721B  &   695 &  3.626 &    &  090530   &   1374 &  1.3 &    &  \textbf{050826}   &   2591 &  0.297  & \\
070802   &   706 &  2.45 &    &  090618   &   1392 &  0.54 &    &    &    &    & \\
070810A  &   714 &  2.17 &    &  090715B  &   1419 &  3 &    &    &    &    & \\
071003   &   767 &  1.6043 &    &  090726   &   1430 &  2.71 &    &    &    &    & \\
071010A  &   774 &  0.98 &    &  090809   &   1443 &  2.737 &    &    &    &    & \\
071010B  &   774 &  0.947 &    &  090812   &   1446 &  2.452 &    &    &    &    & \\
071020   &   784 &  2.145 &    &  090814A  &   1448 &  0.696 &    &    &    &    & \\
071031   &   795 &  2.692 &    &  090926B  &   1490 &  1.24 &    &    &    &    & \\
071112C  &   806 &  0.823 &    &  090927   &   1491 &  1.37 &    &    &    &    & \\
\hline
\label{table_pehdata}
\end{tabular}
\end{minipage}
\end{table*}


\end{document}